# FRET-Enhanced Singlet Fission in Pentacene Derivatives

Oskar Kefer,[a,d] Philipp Ludwig,[b] Benedikt Dittmar,[b] Felix Deschler,[a,d] Jan Freudenberg,*[b] Andreas Dreuw,*[c] Uwe H.F. Bunz*[b,d] and Tiago Buckup*[a,d]

[a] Physikalisch-Chemisches Institut, Ruprecht-Karls-Universität Heidelberg, Im Neuenheimer Feld 229, 69120 Heidelberg, Germany
[b] Organisch-Chemisches Institut, Ruprecht-Karls-Universität Heidelberg, Im Neuenheimer Feld 270, 69120 Heidelberg, Germany
[c] Interdisciplinary Center for Scientific Computing, Ruprecht-Karls-Universität Heidelberg, Im Neuenheimer Feld 205, 69120 Heidelberg, Germany
[d] Institute for Molecular Systems Engineering and Advanced Materials (IMSEAM), Ruprecht-Karls-Universität Heidelberg, Im Neuenheimer Feld 225, 69120 Heidelberg, Germany

* Corresponding Authors: tiago.buckup@pci.uni-heidelberg.de, uwe.bunz@oci.uni-heidelberg.de, dreuw@uni-heidelberg.de, freudenberg@oci.uni-heidelberg.de

Supporting information for this article is given via a link at the end of the document.

**Abstract:** Conversion of solar energy with high quantum efficiencies is a key challenge in energy technologies. Excitation energy transfer (EET) mechanisms, such as Förster resonance energy transfer (FRET), play a crucial role in facilitating minimal energy loss in biological light-harvesting systems by directing excitation energy to conversion centers. Inspired by this, we show that singlet fission (SF) sensitizers are multi-exciton generation centers, to which surrounding molecules funnel excitation energy via FRET. We study the impact of such EET on SF using two structurally distinct yet optically similar pentacene derivatives: a stability-enhanced "Geländer" pentacene, and the well-studied TIPS-pentacene. Transient absorption spectroscopy reveals a $R^{-6}$ dependence of the SF rate on molecular separation R in binary acene:polymethylmetacrylate thin film blends, which is typical for FRET. Optimizing FRET is a promising direction for future improvements in light harvesting using SF materials, inspired by natural light-harvesting complexes.

We herein demonstrate the potential of accessory light-harvesting chromophores to enhance singlet fission by Förster resonant energy transfer towards the molecular singlet fission pair of pentacene derivatives.

Non-radiative excitation energy transfer (EET) allows excitation energy to move loss-less in any type of system. Prominent examples are light-harvesting complexes, where optimally arranged networks of light antennae funnel solar energy towards the reaction center, in which it is converted into chemical energy with near unity quantum efficiency.[2] Beyond biological systems, EET has found widespread applications in (bio-)sensors,[3-5] in conjunction with (organic) semiconductors[6-7] and quantum dots,[8] and has led to breakthrough advances in microscopy.[9] The efficiency of the underlying non-radiative EET process is fundamentally related to the electronic coupling ($V_{el}$) between an energy donor (D) and an energy acceptor (A)

$$V_{el} = \langle DA^*|\hat{H}|D^*A\rangle = V_{coul} + V_{short}, \quad (1)$$

based on Fermi's Golden Rule.[10-11] The coupling can be divided into: a Coulombic coupling term ($V_{coul}$) that is present at all length scales, and a short-range term ($V_{short}$) that depends on the wavefunction overlap. $V_{short}$ can dominate at short distances ($R < 1$ nm) and is described by Dexter EET.[12] The rate of Dexter EET ($k_{DT}$) decays exponentially[13] with intermolecular separation (R) divided by the sum of the van-der Waals radii (L) of the interacting molecules

$$k_{DT} \propto |V_{short}|^2 \propto e^{\frac{-2 \cdot R}{L}}. \quad (2)$$

The van-der Waals radii are usually quite small ($L \ll 1$ nm), leading to a prompt drop-off of $V_{short}$ with increasing R, leaving $V_{coul}$ as primary coupling term at larger separations ($R > 1$ nm). It is responsible for an EET mechanism that is described as Förster resonance energy transfer (FRET),[14] which is dominated by the interaction between the transition dipoles of D and A. The rate of FRET ($k_{FRET}$)[10, 15] scales with $R^{-6}$, according to

$$k_{FRET} = \frac{1}{\tau_{S_1}} * \left(\frac{R_0}{R}\right)^6 \quad (3)$$

and is related to the lifetime ($\tau_{S_1}$) of the donors' emissive state at infinite separation and the Förster-Radius ($R_0$, in nm), defined as

$$R_0 = 0.0211 \cdot \left(\frac{\kappa^2 \cdot \theta_{em} \cdot J}{n^4}\right)^{\frac{1}{6}}. \quad (4)$$

In turn, parameters, such as the relative orientation ($\kappa$) of chromophores, the quantum yield (QY) of emission ($\theta_{em}$) of D, the refractive index (n) and the spectral overlap integral of emission and absorption (J) modulate the rate and efficiency of FRET. Optimizing these parameters can lead to high efficiencies, mimicking the near unity QY observed for the conversion of sun energy in biological systems.[2, 16]

Loss-less ET should be applicable in singlet fission (SF),[17-18] increasing solar conversion efficiency through multi-exciton generation.[19-26] It proceeds via the following mechanism:

$$S_1 + S_0 \leftrightharpoons {}^x[T_1T_1] \to [T_1 \cdots T_1] \to 2 \times T_1 \quad (5)$$

Efficient SF is characterized by rapid conversion of the photo-excited singlet ($S_1$) to the correlated triplet-pair state (${}^x[T_1T_1]$, x =1,3,5)[27] upon interaction with a ground-state chromophore ($S_0$). Separation of ${}^x[T_1T_1]$ into individual triplets ($T_1$) is necessary to extract the nascent triplets.[28] The bi-molecular nature of SF's initial step requires the availability of suitable interaction partners with the photo-excited chromophore. For example, in solution this is decided by the number of encounters (collisions) between molecules due to diffusion,[29-30] following a linear trend of the SF rate with chromophore concentration.[31-32] Effects, such as molecular aggregation,[33-34] are possible causes for deviations from linearity at higher concentrations. In the extreme case, i.e. the bulk material, suitable interaction partners become more abundant resulting in SF rates surpassing even (100 fs)$^{-1}$.[35-36] Here, SF is no longer limited by molecular diffusion, as their motion is frozen, but is governed by the interaction between individual molecules in the bulk. Similar to the description of Dexter EET (*vide supra*), the interaction strength responsible for SF is modulated by the degree of overlap of the molecular wavefunctions.[37] Therefore, small displacements between chromophores have a profound effect on SF, in which increasing molecular separation can lead to an exponentially fast decrease of the rate of SF (eq. 2), while displacement orthogonal to the molecular axis results in more complex trends.[38-39] Consequently, molecular packing within the bulk material is the most decisive factor for intermolecular SF rates.

Given the strong dependence of SF on intermolecular separation, it is surprising that SF can proceed rapidly (< 10 ps) even under non-optimal conditions, such as in amorphous films or diluted blends, where molecules are randomly oriented and farther apart than in optimally packed polymorphs.[40-42] This suggests that an additional EET mechanism may facilitate SF by allowing the singlets to migrate within the bulk until they encounter an appropriate chromophore pair. Following this description, we draw parallels to light-harvesting complexes, where "matrix-molecules" act as light antennae, funneling the excitation energy into SF-capable chromophore pairs, which act as multi-exciton generating reaction centers. Such an idea was presented by Guldi et al.[18, 43], where they presented intramolecular FRET in an oligomeric system with a light harvesting chromophore (subphthalocyanine) covalently bound to a pentacene dimer.

However, little effort has been made to quantify such an aspect in the excited-state dynamics of SF in bulk material to explore its applicability. Our present study aims to address this gap by investigating the role of EET in SF materials using two chromophores: TIPS-pentacene (**TIPS-Pen**) and a novel pentacene derivative, Geländer pentacene[1] (**G-Pen**, Figure 1a). Both molecules exhibit similar absorption (vide infra) and emission profiles,[1] but **G-Pen** additionally features aryl side groups that encapsulate its acene backbone. This drastically enhances **G-Pen**'s stability under ambient conditions while maintaining good solubility, making it an attractive structural motif for organic semiconductor materials and especially spectroscopic studies due to drastically slowed degradation.

Similar to TIPS-ethynylation in **TIPS-Pen** (SI Fig. S1), the encapsulating aryl side groups also decrease the likelihood of strong electronic coupling of the acenes by enforcing a larger separation. This can be seen in the absorption spectra of **G-Pen** after subjecting it to different environments (Figure 1b). Under dilute conditions (0.1 mM, THF), **G-Pen** features a typical spectrum for acenes,[44] sharing remarkable similarities to **TIPS-Pen**, with an absorption maximum at 638 nm (641 nm) and a similar vibronic progression (SI Fig. S1). The spectrum obtained from an amorphous, drop-cast (DC) film appears with only a minor (~4 nm) spectral shift and a weak low-energy tail, indicative of small intermolecular interaction. This stands in stark contrast to other films of acenes ,[35, 44-45] where large spectral shifts and modulations are usually present. Consequently, the diluted blends of **G-Pen** with the host material poly(methylmethacrylate) (PMMA) are expected to be mostly free of aggregates, as confirmed by the nearly unchanged absorption spectra for mass fractions (m%) of 10-20% compared to the spectrum in THF.

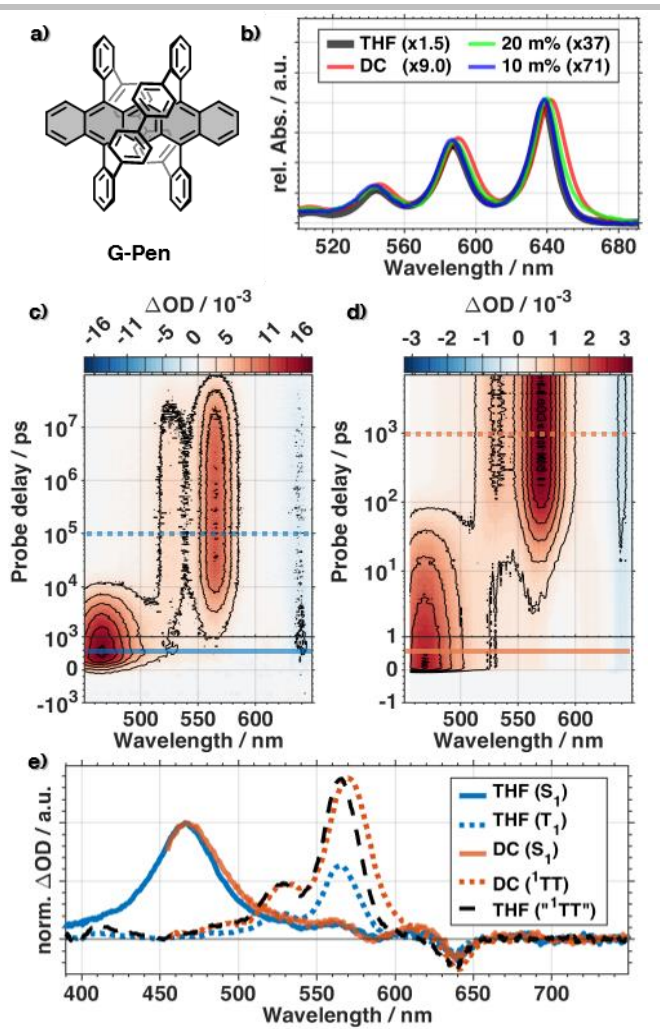

**Figure 1.** a) Molecular structure of racemic **G-Pen** (one enantiomer depicted). b) Absorption spectra of **G-Pen** measured in solution (THF), amorphous, drop-cast (DC) thin film and in blends with PMMA (with mass fractions m% = 10-20%). Spectra are normalized by multiplication with factors given in the legend (see also SI Fig. S1). c) Temporal evolution of transient absorption spectra in solution (THF, up to 100 μs, ca. 300 ps time resolution) and d) as thin film (DC, up to 8 ns, 120 fs time resolution). e) Singlet- and triplet-related spectra (normalized to amplitude of $S_1$). Spectra are taken at time points indicated by horizontal lines in c) and d). The spectrum THF("$^1$TT") is generated by multiplication of the THF($T_1$)-spectrum with a factor of (2/0.92) to account for 2×$T_1$'s in $^1$TT (eq. 5) and reduced ISC due to fluorescence (see Table 1).

To confirm the occurrence of intermolecular SF in **G-Pen**, we compare its transient absorption spectrum under two extreme conditions: highly diluted in THF (Figure 1c) and as an amorphous neat film (Figure 1d). While both initially feature an excited-state absorption (ESA) centered around 460 nm, assigned to the ($S_n \leftarrow S_1$)–transition, the difference of their excited-state dynamics becomes immediately apparent. In dilute THF, concentration-independent (< 1.6 mM, SI section 4) intersystem crossing (ISC) dominates, giving rise to the triplet state ($T_1$) within ca. 5 ns (Table 1), recognized by the ESA at around 560 nm, related to the ($T_n \leftarrow T_1$)–transition. In contrast, as neat film, **G-Pen** exhibits an accelerated transition of $S_1$ to the triplet-related ESA by two orders of magnitude (48 ± 2 ps, SI Table S3), indicative of the bi-molecular SF process. By comparison of the relative spectral profiles of $S_1$ and triplet-related signals (Fig. 1e), we can confirm SF as the primary triplet-formation pathway in the amorphous film, with $^1[T_1T_1]$ ($^1TT$ for brevity) initially produced in a nearly quantitative manner (200% QY for "$T_1$'s").

Moving to intermediate conditions (Figure 2), we study the excited-state dynamics of **G-Pen** diluted in blends of PMMA at different mass fractions, which will be interchangeably used as proxy for the concentration of **G-Pen** ($c_{GP}$) in the films (see SI section 3). As expected for the occurrence of SF, faster $S_1$-decay (Fig. 2a) is observed with increasing $c_{GP}$. Moreover, the maximum relative triplet yield also depends on $c_{GP}$ (combination of $T_1 + {}^xTT + [T_1 \cdots T_1]$, see SI section 5), consistent with bimolecular SF. However, the apparent experimental SF-rate constants ($k_{exp}$) show a non-linear dependence on $c_{GP}$ (Figure 2b, summarized in SI Table S3), deviating from the expected linear trend for bimolecular reactions.[46] Since molecular motion is frozen in PMMA, classical collision theory,[46-47] which predicts a linear trend, does not apply here. Also, models that describe singlet diffusion based on the Smoluchowski rate,[46] which was used by Schmidt et al.[34] to explain the non-linear concentration SF-dependence in solution and to model the dynamics of the SF-reaction of **TIPS-Pen** in blends of PMMA[40] does not apply here either. Instead, alternative explanations for the quadratic concentration dependence ($c_{GP}^2$) of the SF rate must be considered. One explanation involves formation of aggregates at higher concentrations[33] that would also result in a quadratic dependence on the concentration of sensitizers. However, significant amounts of aggregation would result in larger discrepancies of ground- and excited-state spectra at the different concentrations (Figure 1b-e), which we do not observe (*vide supra*). Moreover, we observe ever increasing triplet formation with higher concentrations of **G-Pen**, resulting from the generation of $^1TT$ over $T_1$, with nearly 200% QY for the triplets in blends where $c_{GP} > 250$ mM (see SI section 5).

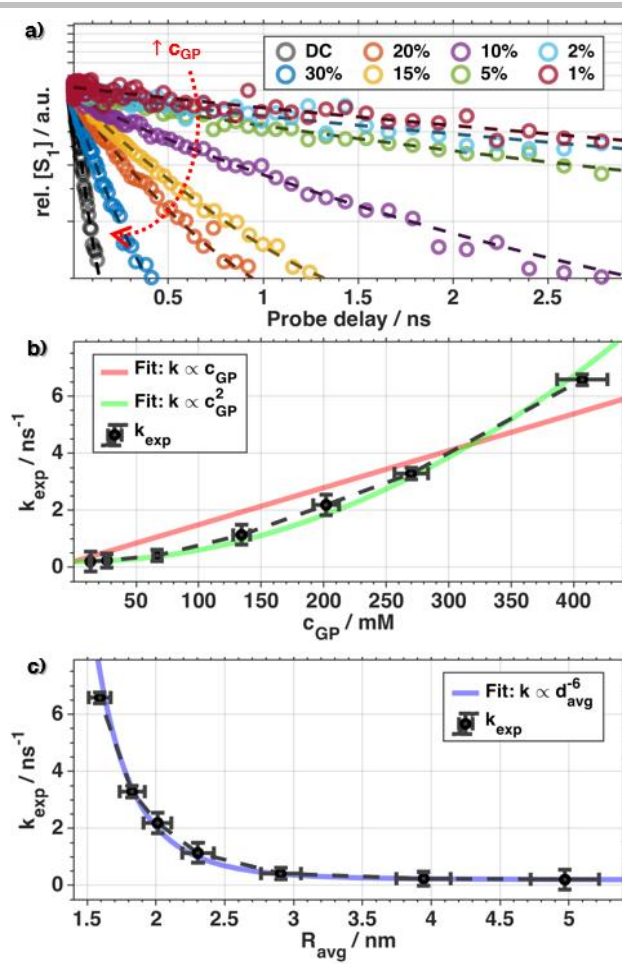

**Figure 2.** Concentration dependence of **G-Pen** in PMMA blends. a) Time-resolved traces averaged over over a spectral window of (440±5) nm, displaying the relaxation kinetics of $S_1$. Experimental rates ($k_{exp}$) obtained from global analysis of transient absorption data (see SI for further details) in the different films are plotted with concentration ($c_{GP}$) in b) and with average separation of chromophores ($R_{avg}$) in c). A linear and quadratic fit (plus offset: (plus offset: (5.06 ns)$^{-1}$, Table 1) to the data is also shown in b). The fit function in c) is given by eq. 3 (plus offset), corresponding to the $R_{avg}^{-6}$-dependence mentioned in the legend.

Consequently, we can rule out aggregation as an explanation for the observed trend (Figure 2b). Even if small amounts of aggregates were present even at high concentrations, they either do not hinder SF or are too few and too separated to significantly impact the overall $^1TT$-yield (*vide supra*).

Instead, the observed "pseudo"-quadratic dependence on $c_{GP}$ can be understood by shifting our perspective to the molecular arrangement within the blend at different concentrations. From the concentration, we calculate the average molecular separation ($R_{avg}$) (see SI section 3 for details), which represents an average over the entire bulk. Plotting experimental rates against $R_{avg}$ (Fig. 2c), we find that a $R_{avg}^{-6}$–dependence, which suggests that FRET (eq. 3), describes our data best (fitted with $k_{exp} = k_{FRET} + (\tau_{S_1})^{-1}$, see also Table 1). FRET explains the occurrence of the "pseudo-$c_{GP}^2$" relationship, which arises naturally from the $R_{avg}^{-6}$–trend ($c_{GP}^2 \propto R_{avg}^{-6}$, see also SI section 3). However, at a first glance, it is surprising that the SF rate follows a FRET formalism, considering the underlying molecular interaction for the two processes are not compatible: While transition dipole-transition dipole interaction is responsible for the occurrence of FRET, SF is modulated by the overlap integral of the wavefunctions, which follows an exponential "Dexter"-like distance scaling (eq. 2). Fitting the experimental data shown in Fig. 2c accordingly, however, fails to describe the dependence on the average molecular separation $R_{avg}$ (see SI section 6). Moreover, the agreement with the FRET formalism (eq. 3) is so robust that individual points of the experimentally determined rates can be fitted individually to predict the remaining rate constants – something that is not possible assuming exponential dependence on $R_{avg}$. This result is in agreement with the fact that Dexter-EET is only relevant for intermolecular distances below 1 nm and here the average separation of the molecules is well above 1 nm. Thus, the most plausible explanation for why the SF rate follows a distance scaling according to a transition dipole-transition dipole interaction is that we are not observing the intrinsic SF rate of **G-Pen** (or **TIPS-Pen**), but a FRET process as preceding rate-limiting step; The picture emerges that singlets diffuse within the bulk material via FRET until they arrive at a suitable molecule pair with a fitting arrangement for SF.

**Table 1.** Parameters used for calculation of FRET-rates according to eq. 3 (see Figure 3). $\tau_{S_1}$(**G-Pen**) is determined from the fit of transient absorption spectra measured in THF (< 1.6 mM, SI section 4). $R_0^{exp}$ is determined by fitting the experimental rate constants in Figure 2 with $k_{exp} = k_{FRET} + (\tau_{S_1})^{-1}$. Spectra for the calculation of J were taken from Ref. [1] (see also SI section 7). $\kappa^2$ was set constant to a value of 0.476, according to Ref. [15] and refractive index was calculated as a weighted average of n(PMMA)[48] and n(Molecule) (details are provided in SI section 7).

|  | **G-Pen** | **TIPS-Pen** |
|---|---|---|
| $S_1$-Lifetime $\tau_{S_1}$ [ns] | 5.06 ± 0.05 | 12[34, 49] |
| Emission Quantum Yield $\theta_{em}$ [%] | 8[1] | 44[1] |
| Spectral Overlap Integral **J** [mol$^{-1}$ L cm$^{-1}$ nm$^4$] | 0.80×10$^{15}$ | 2.24×10$^{15}$ |
| *Calculated* FRET-Distance $R_0^{calc}$ [nm] | 2.83 | 4.42 |
| *Experimental* FRET-Distance $R_0^{exp}$ [nm] | 2.88 ± 0.14 | 4.52 ± 0.25 |

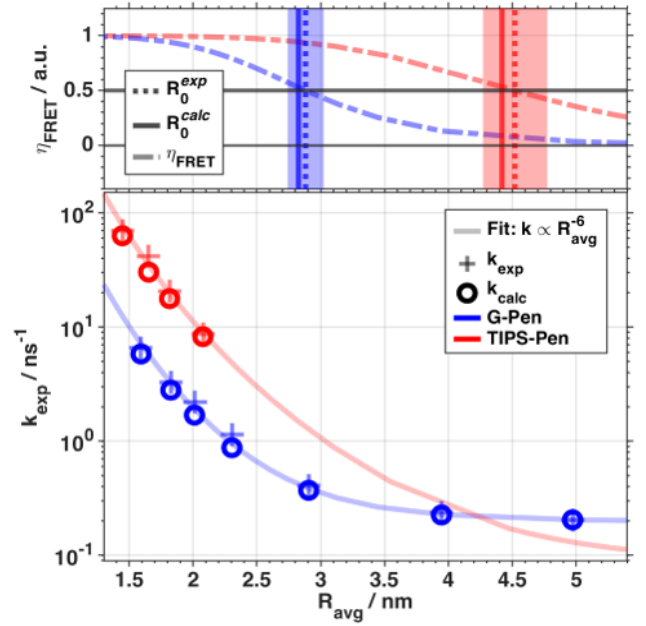

**Figure 3.** Comparison of experimental results with FRET-theory. The top panel displays the FRET efficiency ($\eta_{FRET}$, eq. 6), determined from the Förster-Radius $R_0$ (eq. 4). The experimental Förster-Radius ($R_0^{exp}$) is determined from the fit of experimental rates with eq. 3 (plus offset). The shaded area around $R_0^{exp}$ show the 95% confidence interval of the fitted values (standard deviation). Förster-Radius values according to theory ($R_0^{calc}$) are calculated [15] using eq. 3 and 4 with parameters given in Table 1 (more details can be found in SI section 7). Experimental- ($k_{exp}$) and calculated rates ($k_{calc} = k_{FRET}^{calc} + (\tau_{S_1})^{-1}$) with the corresponding fit of experimental data is shown in the bottom panel. Individual rate constants of **G-Pen** are shown in blue and of **TIPS-Pen** in red and can be found in SI Table S5.

We can further scrutinize this proposed mechanism by calculating experimental rates (according to eq. 3 and 4), based on results from this study and literature precedence (summarized in Table 1, see also SI section 7 for more details). A comparison of calculated ($k_{calc}$) and experimental ($k_{exp}$) rates (Figure 3) corroborates our hypothesis that FRET precedes SF, as it also accounts for the observed differences of the macroscopic "SF"-rates (FRET rate-limited) of **TIPS-Pen** in comparison to **G-Pen**, explained entirely by the formalism for FRET.

As the rate and FRET-efficiency ($\eta_{FRET}$, Figure 3), with

$$\eta_{FRET} = \frac{R_0^6}{R_0^6 + R_{avg}^6} \qquad (6)$$

depend only on $R_0$ (eq. 4), it is now straightforward to extrapolate which parameters (summarized in Table 1) lead to good materials for SF applications exploiting FRET for light harvesting: (i) A large overlap between donor- and acceptor spectra, and (ii) a high QY of emission. While **G-Pen** may not be optimal in this regard in comparison to **TIPS-Pen** (8% vs 44%), a blend with another, more emissive material (e.g. TIPS-tetracene, $\theta_{em}$ = 79%)[50] in conjunction with **G-Pen** (or even better, a dimer capable of intra-molecular SF) might show enhanced SF characteristics compared to blends with pristine **G-Pen** by optimizing the critical factors (*vide supra*) of efficient FRET. Such a system would be comparable to the antennae compounds found within light-harvesting complexes, where the excitation energy absorbed by the donors in the blend would be funneled towards reaction centers (in this case, SF-capable dimers). **G-Pen**, with its fivefold increased triplet-lifetimes in comparison to **TIPS-Pen** (50 μs versus 10 μs, see SI Tables S3-4) would provide a high likelihood of triplet-diffusion to generate higher quantities of excitons that can be harvested.

In conclusion, we demonstrated efficient singlet fission in a novel pentacene derivative **G-Pen**, depending on concentration. A "pseudo"-second order concentration dependence of the singlet fission rate of **G-Pen** in PMMA blends has been revealed to be fundamentally caused by transition dipole-transition dipole interactions between singlets. The accelerated SF kinetics at high concentrations are explained by FRET that precedes SF and in turn, acts as its rate-limiting step. The approximately 10x faster kinetics in **TIPS-Pen** can be rationalized by its improved overlap integral of absorption- and emission- spectra and its enhanced fluorescence quantum yield. These findings build a foundation for further applications for SF, where, instead of neat films that potentially bring disadvantageous reductions in triplet lifetimes, blends of SF chromophores with "light-harvesting" chromophores displaying improved FRET capabilities should be considered, drawing inspiration from biological light-harvesting complexes.

## Supporting Information

Supplementary Information that supports the findings presented here can be found free of charge, appended to the document. SI contains: Experimental section, UV-Vis- and TA spectra and fit results, and details regarding the calculation of FRET rates. The authors have cited following additional references within the Supporting Information: [51-56]

## Acknowledgements

Authors thank the Deutsche Fördergemeinschaft for funding this study through the Sonderforschungsbereich (SFB) 1249 projects B01, B04, B09 and A01.

**Keywords:** singlet fission, excitation energy transfer, FRET, kinetics, time-resolved spectroscopy, reaction mechanisms

# Supplementary Information for

# FRET-Enhanced Singlet Fission in Pentacene Derivatives

Oskar Kefer,[a,d] Philipp Ludwig,[b] Benedikt Dittmar,[b] Felix Deschler,[a,d] Jan Freudenberg,*[b] Andreas Dreuw,*[c] Uwe H.F. Bunz*[b,d] and Tiago Buckup*[a,d]

[a] Physikalisch-Chemisches Institut, Ruprecht-Karls-Universität Heidelberg, Im Neuenheimer Feld 229, 69120 Heidelberg, Germany

[b] Organisch-Chemisches Institut, Ruprecht-Karls-Universität Heidelberg, Im Neuenheimer Feld 270, 69120 Heidelberg, Germany

[c] Interdisciplinary Center for Scientific Computing, Ruprecht-Karls-Universität Heidelberg, Im Neuenheimer Feld 205, 69120 Heidelberg, Germany

[d] Institute for Molecular Systems Engineering and Advanced Materials (IMSEAM), Ruprecht-Karls-Universität Heidelberg, Im Neuenheimer Feld 225, 69120 Heidelberg, Germany

* Corresponding Authors: tiago.buckup@pci.uni-heidelberg.de, uwe.bunz@oci.uni-heidelberg.de, dreuw@uni-heidelberg.de, jan.freudenberg@oci.uni-heidelberg.de

# Table of Content

# 1. Experimental Section

### Sample preparation

Pentadecacyclo[48.6.6.2⁸,¹¹.2¹²,¹⁵.2³⁶,³⁹.2⁴⁰,⁴³.0²,⁷.0¹⁶,²¹.0²²,⁵³.0²³,²⁸.0²⁹,⁵⁴.0³⁰,³⁵.0⁴⁴,⁴⁹.0⁵¹,⁵⁶.0⁵⁷,⁶²]hepta-conta-1(56),2,4,6,8,10,12,14,16,18,20,22,24,26,28,30,32,34,36,38,40,42,44,46,48,50,52,54,57,59,61,63,65,67,69-pentatriacontaene ("Geländer" pentacene, **G-Pen**) was synthesized according to the protocol reported in ref. [1], yielding a racemic mixture (one enantiomer shown in Fig. 1 in the main text). 6,13-Bis((tri*iso*propylsilyl)ethynyl) pentacene (**TIPS-Pen**) was synthesized according to the protocol reported in ref. [52]. Poly(methyl methacrylate) (PMMA) (average $M_w \approx 15000$ g/mol) was purchased directly from Sigma Aldrich and used without further purification.

### Film preparation (blends and thin film)

**G-Pen** or **TIPS-Pen** were co-dissolved with poly(methylmethacrylate) (PMMA) in chloroform with a total concentration of 10 mg/mL and mass fractions (m%) specified in Table S2. Samples were sonicated for 30 min to aid solvation. The solutions were spin-coated onto glass substrates, which were precleaned by sonication in acetone, isopropyl alcohol and ethanol followed by cleaning with oxygen plasma. The films were obtained after spin-coating for 40 s at 1000 rpm with an acceleration of 500 rpm/s under ambient conditions.

### UV-Vis Absorption

Steady-state absorption was measured with Shimadzu UV-1800 and UV-2600 spectrometers. Solution measurements of **G-Pen** were carried out using a stem solution with concentration c = 1.6 mM, which was prepared by dissolving 5.7 mg **G-Pen** (M = 883.1 g·mol$^{-1}$) in 4 mL tetrahydrofuran (THF, ≥99.9%, Sigma Aldrich). The exact concentration was cross-checked by comparison with the extinction coefficient from literature ($\varepsilon$(640 nm, THF) = 11.16·10$^3$ M$^{-1}$·cm$^{-1}$)[1] by measurement of the absorption (A) in a 500 μm cuvette (A(640 nm, THF) = 0.900 OD {expected}, A(640 nm, THF) = 0.904 OD {measured}). Less concentrated THF samples were prepared by diluting part of the stem solution with THF. Their concentrations (Fig. S2) were also corroborated by measuring their absorbance in 1-2 mm cuvettes.

Absorption of solid-state samples (neat film and blends in PMMA) was measured by clamping the glass substrates inside the illuminated area of the spectrometer with a home-built adapter. Absorbance of **G-Pen** in PMMA with a mass fraction of 1 m% was not resolvable and it was used as reference substrate (to internally correct for e.g. scattering, reflection of the glass substrate and deposited PMMA). This sample was put in the reference beam of the spectrometer during absorption measurements of the other samples.

### Transient Absorption

Transient absorption was measured with commercially available experimental setups "Helios" and "EOS" developed by Ultrafast Systems. A description of the full experimental layout can be found in ref. [29]. To summarize briefly, part (45%) of the output of a Ti:Sa laser system (Coherent Astrella, 90 fs, 1.5 mJ, 4 kHz) drives commercially available optical parametric amplifier systems (Coherent TOPAS) to generate laser pulses with variable spectra from ranging from UV-VIS to IR. The excitation spectrum for time-resolved experiments was centered around 640 nm (330±30 cm$^{-1}$ (FWHM), 90±15 fs pulse duration). Continuum white light probe for "Helios" (short-time, up to 8 ns) is generated by focusing a fraction of the fundamental (<1%) inside optical media, such as sapphire (for **G-Pen**) and CaF$_2$ (for **TIPS-Pen**). Probe delay is set by redirecting the probe beam over an opto-mechanical delay stage. Relative polarization of the probe is set to 54° (magic angle) to eliminate anisotropic signal contributions. For "EOS" (long-time, <1 ms (1 kHz)$^{-1}$), an external, pulsed white-light laser source (unpolarized) is used and electronically delayed. Probe delays up to 150 µs were measured, based on the observed signal dynamics. Pump- and probe beams are focused to spots of 300 µm in diameter. Excitation energies were varied, depending on the sample (see Figs. S2, S4, S6, S8, S10, S13 and S15 for details). They were kept low as possible to stay within linear excitation regimes (checked before measurement by comparison of spectra amplitudes at different fluences, *not shown*) and to limit photo-induced damage. Time resolution of short- and long-time measurements was determined by fit of the instantaneous response, resulting in timescales of 120 fs (FWHM) and 300 ps (FWHM), respectively.

Experiments on solutions were carried out in d = 0.5-2 mm cuvettes (d ≤ 1 mm: flow-cells {2 mL/min}, Fig. S2). Cuvettes with d = 2 mm were equipped with a teflon-coated magnetic stirrer that is stirred continuously during measurement. Solid-state samples were continuously (and randomly) moved within a 2x2 mm square to minimize thermal effects, degradation of the sample and to measure homogeneously over the sample.

## 2. Steady-State Absorption

Normalized absorption spectra of **G-Pen** and **TIPS-Pen** in THF are compared in Figure S1a. Other than **TIPS-Pen** featuring smaller relative amplitudes for higher vibronic transitions ($\nu$ = 1, 2 at ca. 590 nm and 545 nm, respectively), the spectra of both compounds share remarkable similarities under these dilute conditions. The resilience of **G-Pen** towards aggregation becomes evident by comparing the spectra measured at high concentrations in the blends with the spectra measured under dilute conditions (Fig. S1b-c). Minute differences can be found in the spectra of **G-Pen** in the blends of PMMA (Fig. S1b) compared to the solution spectra, even at the highest mass fraction (m% = 30%). For **TIPS-Pen**, slightly larger shifts and spectral broadening compared to **G-Pen** can be found, pointing towards increased intermolecular interaction as the concentration increases.

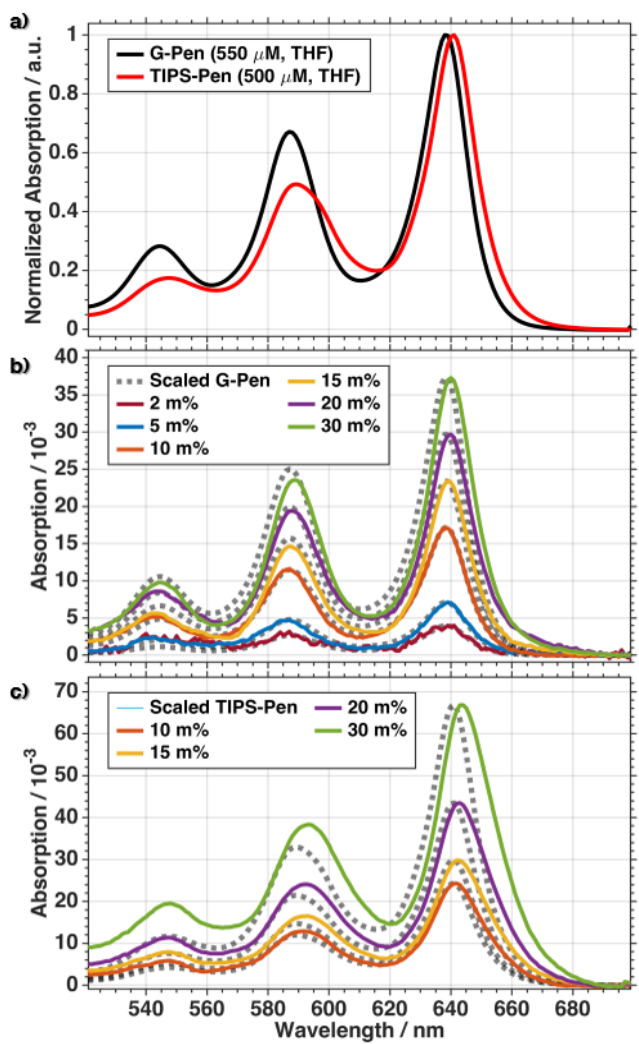

**Figure S4**. Comparison of steady-state absorption spectra of **G-Pen** and **TIPS-Pen**. a) Normalized spectra measured in solution (THF). b) and c) Absorption spectra of PMMA blends of **G-Pen** and **TIPS-Pen**, respectively, with different mass fractions (m%). Grey-dotted lines represent scaled versions of spectra shown in a) to facilitate comparison between spectra.

## 3. Estimation of Concentrations and Intermolecular Distances

The concentration ($c_x$) for **G-Pen** (x = GP) and **TIPS-Pen** (x = TP) was estimated based on their relative mass fractions (m%) in the PMMA-blends, in accordance with e.g. ref. [40]. With parameters (summarized in Table S1), such as the densities ($\rho_x$) of the solid materials and their molar weights ($M_x$), $c_x$ is calculated according to

$$c_x = \frac{N_x^{mol}}{V_x^{tot}} = \frac{(m_x \cdot M_x^{-1})}{\frac{m_{PMMA}}{\rho_{PMMA}} + \frac{m_x}{\rho_x}} = \frac{m\% \cdot M_x^{-1}}{\frac{1 - m\%}{\rho_{PMMA}} + \frac{m\%}{\rho_x}}, \tag{S1}$$

where $N_x^{mol}$ is the molar number of molecules and $V_x^{tot}$ is the total volume they occupy. For the calculation of $c_x$, the weighed-out masses of samples ($m_x$) and PMMA ($m_{PMMA}$) can be replaced by mass fractions considering the total mass ($m_{tot}$) cancels out of the equation after substituting $m_x = m_{tot} * m\%$ and $m_{PMMA} = m_{tot} * (1 - m\%)$. Hence, the concentration is calculated based on the three parameters m%, $M_x$ and $\rho_x$.

**Table S2**. Parameters used for calculation of the concentration $c_x$ and average molecular separation $d_x$. **TIPS-Pen** and **PMMA** parameters were taken directly from Sigma Aldrich (can also be found in e.g. ref. [40]). Parameters for **G-Pen** were taken from ref. [1].

|          | $\rho_x$ / $10^3$ g·L$^{-1}$ | $M_x$ / g·Mol$^{-1}$ |
|----------|------------------------------|----------------------|
| **G-Pen**    | 1.240 | 883.10 |
| **TIPS-Pen** | 1.104 | 639.07 |
| **PMMA**     | 1.188 | -- |

The average intermolecular separation ($R_x$) was calculated based on descriptions presented in refs. [40] and [51]. In short, the volume ($R_x^3$) that each molecule occupies (average over the bulk) can be estimated from $c_x^{-1}$. For the conversion to the average volume for individual molecules, the molar number of molecules ($N_x^{mol} = N_x/N_A$) in eq. S1 is substituted by the number of individual molecules ($N_x$) using Avogadro's number ($N_A$) resulting in

$$R_x^3 = \frac{1}{N_A \cdot c_x}$$
$$R_x = \frac{1}{\sqrt[3]{N_A \cdot c_x}}$$
$$R_x = \sqrt[3]{\frac{\frac{1 - m\%}{\rho_{PMMA}} + \frac{m\%}{\rho_x}}{N_A \cdot m\% \cdot M_x^{-1}}} \tag{S2}$$

The estimated values for concentrations and average intermolecular separation for **G-Pen** ($c_{GP}$, $R_{GP}$) and **TIPS-Pen** ($c_{TP}$, $R_{TP}$), based on eqs. S1-2, are summarized in Table S2.

**Table S3**. Tabulated parameters for mass fractions (m%), concentrations ($c_x$), average intermolecular separations ($R_x$) for **G-Pen** (x = GP) and **TIPS-Pen** (x = TP). $c_x$ and $R_x$ calculated according to eqs. S1 and S2, respectively, using parameters from Table S1. A deviation of 5% for the mass fractions was estimated.

|  | Mass fraction m% / % | 1 ± 0.05 | 2 ± 0.10 | 5 ± 0.25 | 10 ± 0.50 | 15 ± 0.75 | 20 ± 1.0 | 30 ± 1.5 |
|---|---|---|---|---|---|---|---|---|
| **G-Pen** | Concentration $c_{GP}$ / mM | 13.4 ± 0.7 | 26.7 ± 1.3 | 66.9 ± 3.3 | 134 ± 6.7 | 202 ± 10 | 270 ± 13 | 407 ± 20 |
|  | Avg. separation $R_{GP}$ / nm | 4.97 ± 0.25 | 3.95 ± 0.20 | 2.90 ± 0.15 | 2.30 ± 0.12 | 2.01 ± 0.10 | 1.83 ± 0.09 | 1.59 ± 0.08 |
| **TIPS-Pen** | Concentration $c_{TP}$ / mM | -- | -- | -- | 181 ± 9.1 | 269 ± 13 | 355 ± 18 | 524 ± 26 |
|  | Avg. separation $R_{TP}$ / nm | -- | -- | -- | 2.09 ± 0.10 | 1.83 ± 0.09 | 1.67 ± 0.08 | 1.46 ± 0.07 |

# 4. Transient Absorption – G-Pen in THF

The excited-state dynamics of **G-Pen** under dilute conditions in THF are evaluated in the following. Figure S2 shows the time-evolution of transient absorption spectra with concentrations ranging from 0.13-1.6 mM. Essentially the same evolution can be observed in this concentration range. This points towards a concentration-independent relaxation pathway, such as intersystem crossing, to be responsible for the conversion of the initially excited singlet state ($S_1$). Thus, we assign the excited-state absorption (ESA) centered around 460 nm to the absorption of $S_1$ in **G-Pen** (Fig. S2), which transitions to a triplet state ($T_1$), with a characteristic ESA centered around 560 nm. This similarity of excited-state dynamics is further stressed in Fig. S3, where results of the fits with global (target-) analysis are presented. Two time-constants describe the complete excited-state evolution of the $S_1$ towards the $T_1$ and finally back to the electronic ground-state ($S_0$), which is modelled by a sequential evolution of the three states in Fig. S3c. Using this linear kinetic model, we calculate the species-associated difference spectra (SADS) for the different concentrations. By comparing SADS (Fig. S3c), normalized to the amplitude of the $S_1$-ESA, we evaluate the relative quantum yield (QY) for the formation of $T_1$ for different concentrations: The extinction coefficient for a given state is constant, and as result, differences in the amplitudes reflect changes in the transition efficiency. Since there is a near-perfect overlap between the scaled $T_1$-SADS, the QY for ISC in **G-Pen** is constant in the concentration range 0.13-1.6 mM. As we will show later and as shown in the main text (Fig. 1), concentration-dependent dynamics manifests in different relative amplitudes of triplet-related spectra compared to the initial $S_1$-ESAs.

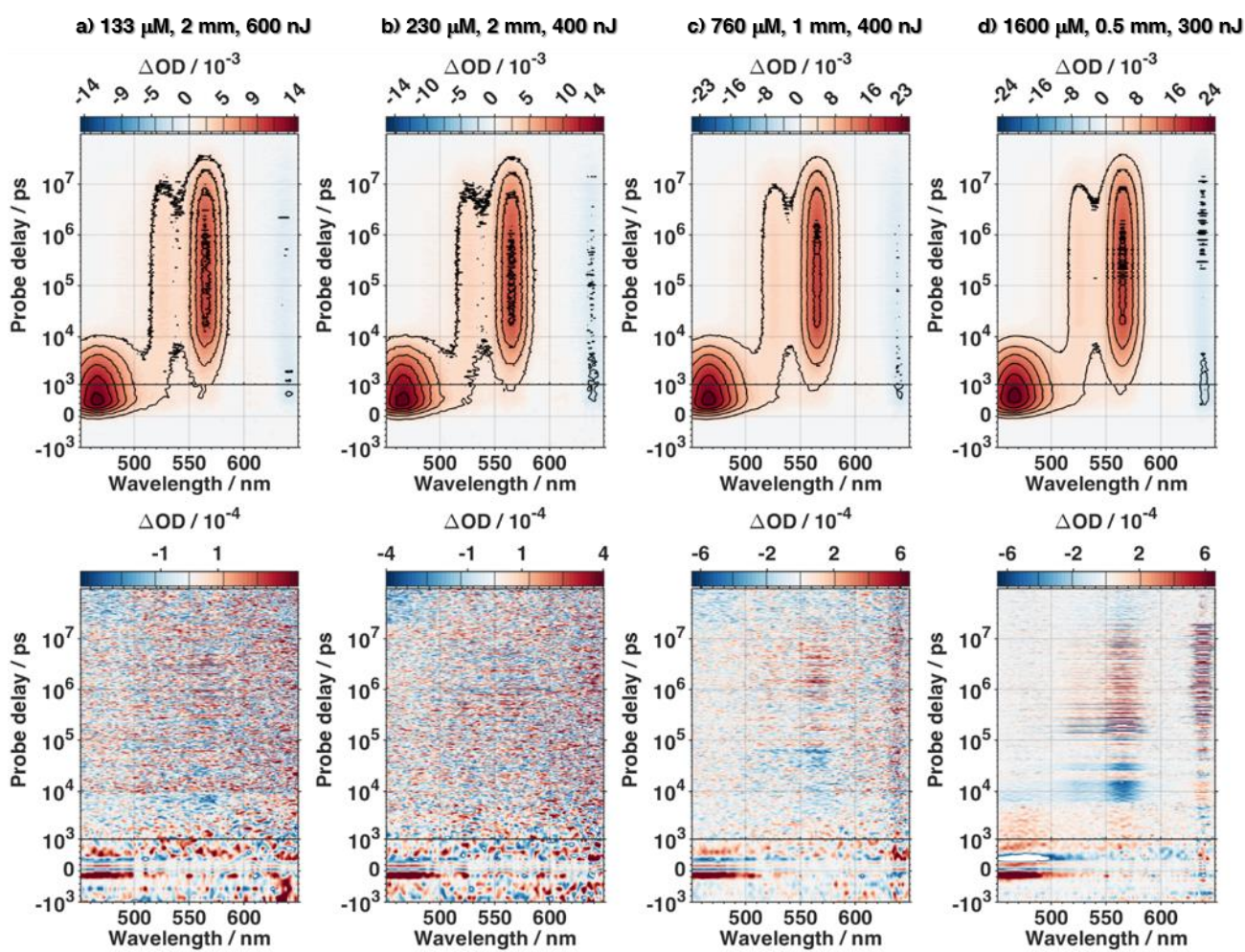

**Figure S5**. Evaluation of temporal evolution of transient absorption spectra of **G-Pen** in THF. Experimental results for different concentrations ranging from 0.13 mM to 1.6 mM are shown in the top panels of a)-d). Black contours (top) are drawn at 15% intervals. The bottom panels show the residuals after subtraction of the fit (global analysis, time constants shown in Fig. S3). The colour scale (bottom) is scaled down to 2.5% of the original data above (notice changed scaling to $10^{-4}$). Details about the individual experiments (concentration, cell width, excitation energy) are given above the figures.

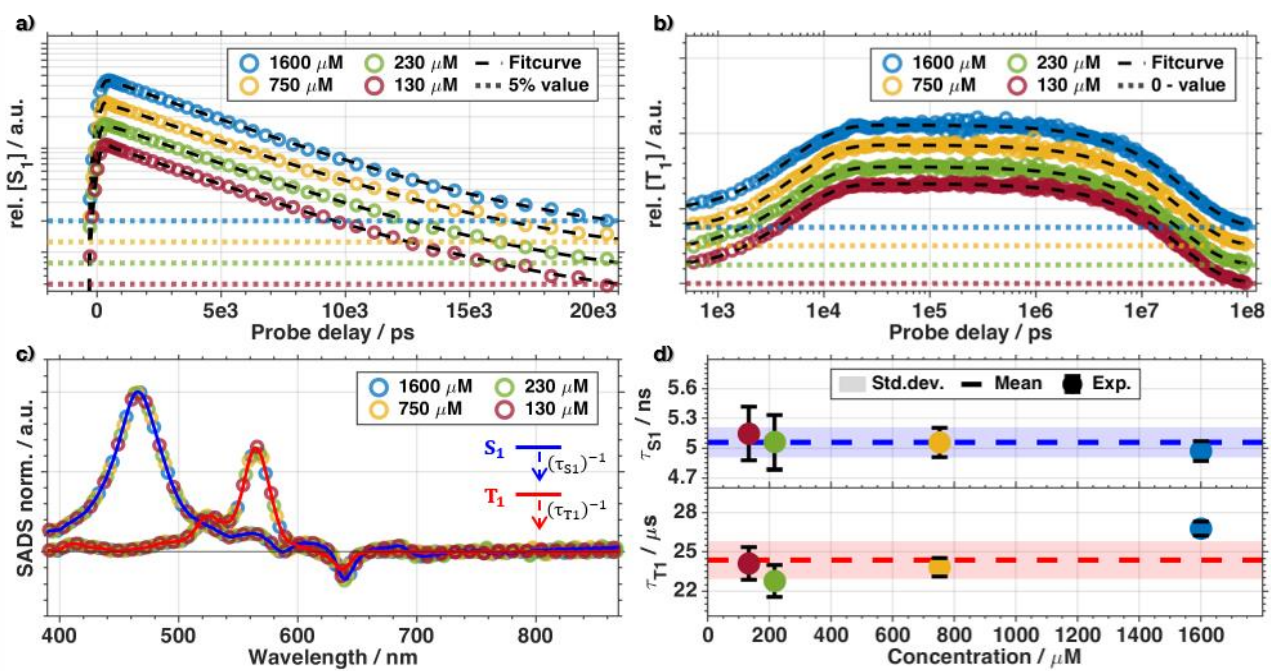

**Figure S6.** Summary of measurements on **G-Pen** in THF under dilute conditions. a) and b) show the kinetic traces related to the $S_1$ (averaged over 460 ± 5 nm) and $T_1$ (averaged over 560 ± 5 nm), respectively, normalized to the maximum amplitudes of the traces in a). For better visual clarity, the amplitudes in the semi-log plot in a) are multiplied by factors ($4^{n/3}$, n = 0,1,2,3). The 5%-threshold of relative signal is indicated by dotted lines in a). In b), the individual traces are shifted up by 0.125*n (n = 0,1,2,3) for more visual clarity. The individual baselines ("0-value") are indicated by the dotted lines. c) Amplitude normalized SADS after fitting the transient spectra (Fig. S1) with a sequential kinetic model shown as an inset. The SADS of the individual measurements are normalized to the amplitude of the $S_1$-ESA at 460 nm. For more visual clarity, 50 data points out of the total 800 data points collected by the spectrometer are shown. The selected points are shifted for the individual measurements to reflect their similarity. As visual aid, averaged spectra (red and blue lines) are also shown. c) Time constants obtained from global analysis of the datasets shown in Fig. S1. Averaged results and standard deviations (Table 1 in main text and Table S3) are indicated by dashed line and coloured area, respectively.

# 5. Transient Absorption on Solid-State Samples

The solid-state samples are evaluated next. We will start by presenting the experimental results of **G-Pen** in Figs. S4-11, recorded with the "short-time, Helios" and the "long-time, EOS" TA setups. Figure S12 will summarize the main findings relevant to the findings presented in the main text. Similarly, experimental results and a summary of the analysis for **TIPS-Pen** are presented in Figs. S13-16 and Fig. S17, respectively. Most relevant time constants are also summarized in Tables S3 and S4 for **G-Pen** and **TIPS-Pen**, respectively. Due to increased scattering contributions in "long-time, EOS" measurements (due to considerably weaker white-light intensities), the spectra for wavelengths greater than 600 nm were omitted. Since no meaningful signal is observed in the range of 650-800 nm (see Fig. S3c), the quality of the fit is not altered by the omission of data in this spectral window.

In general, both films show multiple time constants (<5000 ps in **G-Pen** (Figs. S5 and S7) and <500 ps in **TIPS-Pen** (Fig. S14)) that are related to the decay of the initial excited-state singlet ($S_1$), hinting at a distribution of different intermolecular separations, as would be expected for the blends in PMMA. However, from those decay-associated difference spectra (DADS), it also becomes clear that there is one dominant time scale, during which most of the singlet spectrum is converted to the corresponding triplet spectrum. They are summarized in Tables S3 and S4 for **G-Pen** and **TIPS-Pen**, respectively. These time constants describe the rate-limited step of FRET on SF for the averaged intermolecular separations (Table S2), which are calculated over the whole bulk and represent the majority portion of the distribution of relative intermolecular distances. Moreover, it is to be noted that for **TIPS-Pen,** one time constant (<3 ps), which is significantly shorter than the FRET-rate limited step, can be observed at all measured mass fractions (10-30%), despite the majority of singlets decaying during the time frame described by the subsequent time-constant (see also Table S4). This observation is less pronounced in **G-Pen**, which is very consistent with the relative spectral shifts observed between PMMA blends and the diluted THF samples, which may point toward more significant fraction of aggregation of **TIPS-Pen** in the blends compared to **G-Pen**. This further strengthens our assumption that aggregates play a minor role in the observed "pseudo"-$c^2$ trend of the experimental rate constants (Figure 2 in the main text, $\tau_{S_1 \to "T"}$ in Table S3-4).

## G-Pen – Experimental Data and Fits

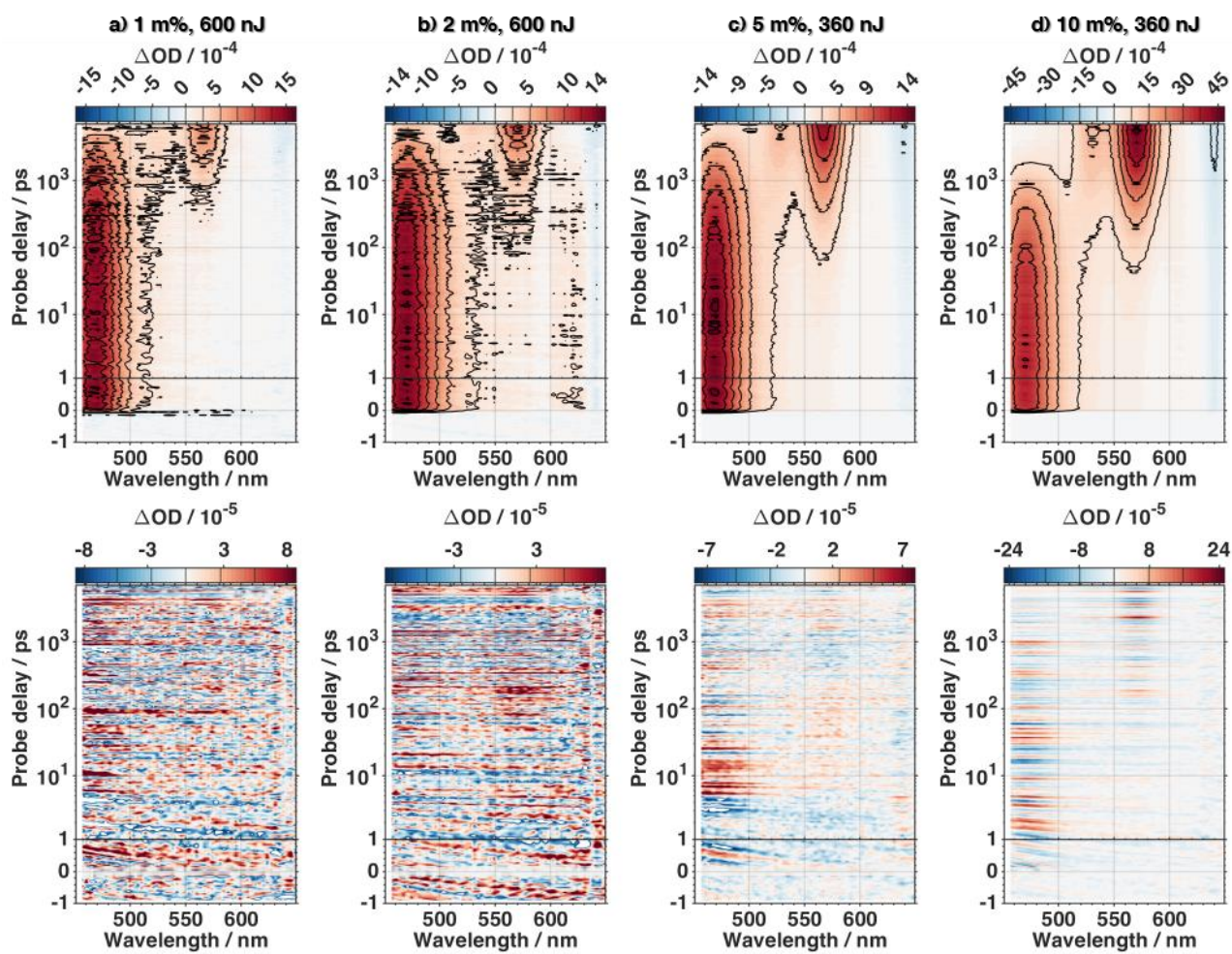

**Figure S7**. Evaluation of temporal evolution of transient absorption spectra of **G-Pen** in blends of PMMA (1-10 m%) measured with "short-time, Helios" TA setup. Experimental results for different mass fractions are shown in the top panels of a)-d). Black contours (top) are drawn at 15% intervals. The bottom panels show the residuals after subtraction of the fit (global analysis, time constants can be found in Fig. S5 and Table S3). The colour scale (bottom) is scaled down to 5% of the original data above (notice changed scaling to $10^{-5}$). Details about the individual experiments (mass fraction, excitation energy) are given above the figures.

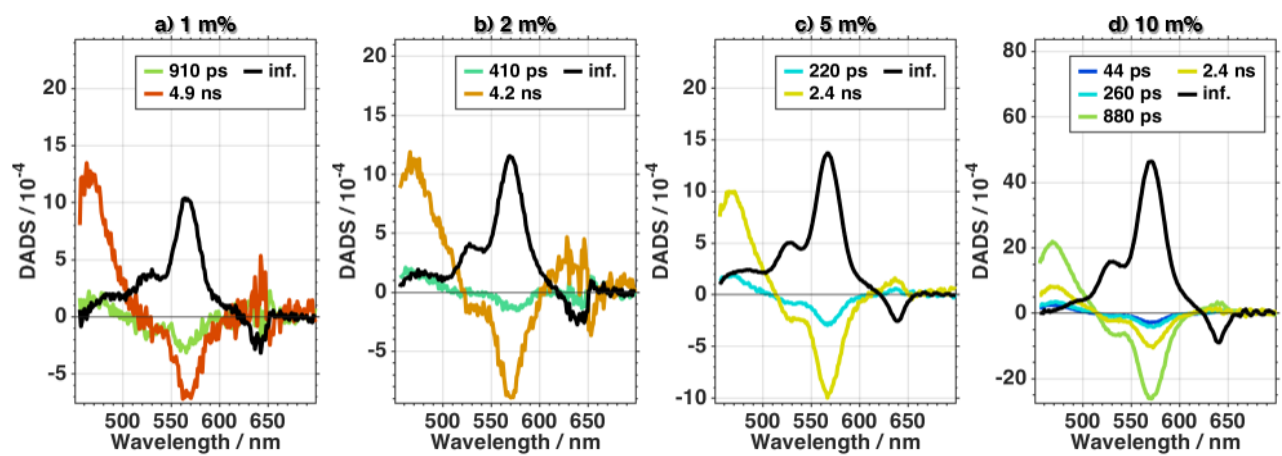

**Figure S8**. Decay-associated difference spectra (DADS) generated from the global analysis of results on **G-Pen** acquired with "short-time, Helios" TA setup. a)-d) DADS for experimental data shown in Figure S4.

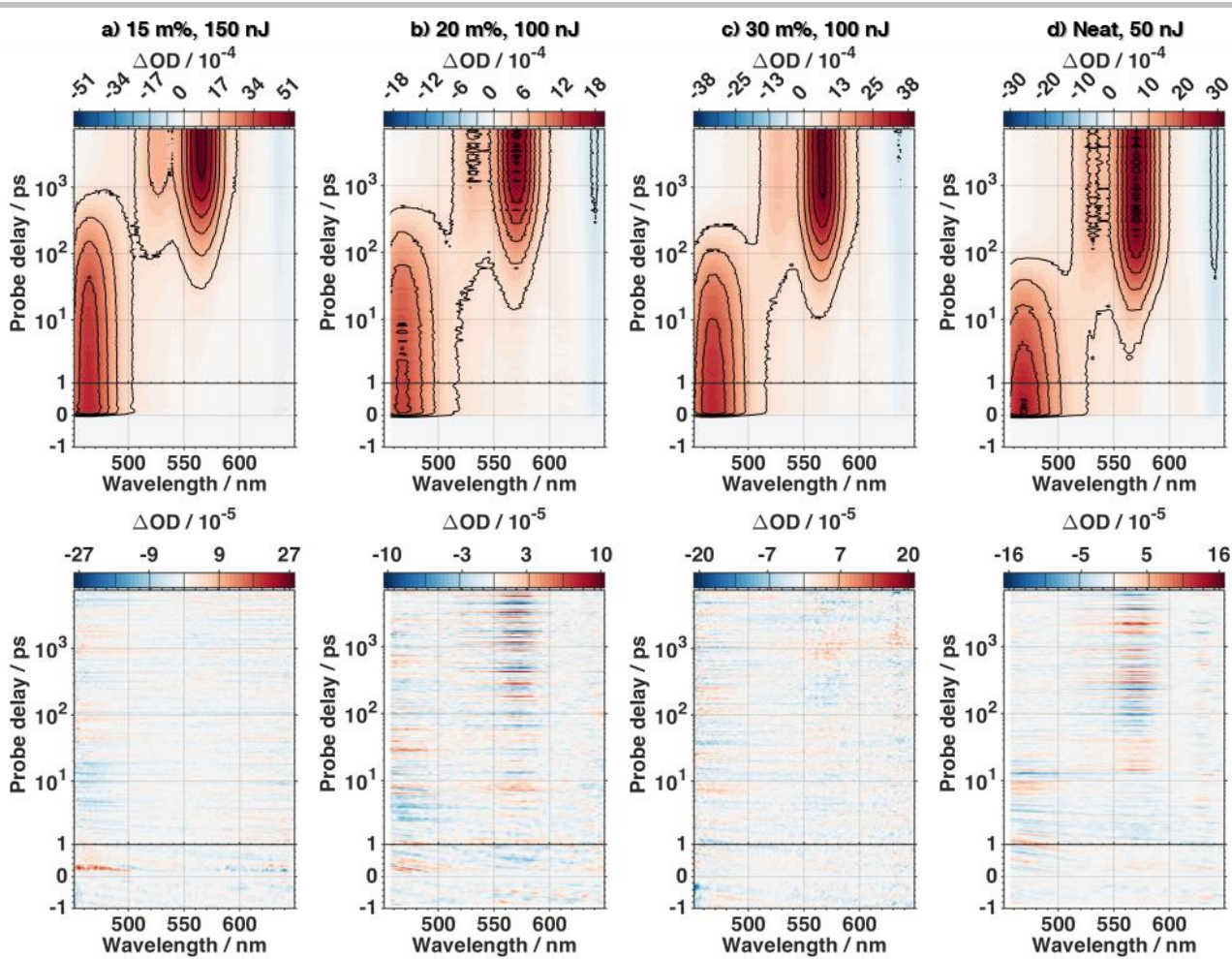

**Figure S9**. Evaluation of temporal evolution of transient absorption spectra of **G-Pen** in blends of PMMA (10-30 m%) and neat film measured with "short-time, Helios" TA setup. Experimental results for different mass fractions and the neat film are shown in the top panels of a)-d). Black contours (top) are drawn at 15% intervals. The bottom panels show the residuals after subtraction of the fit (global analysis, time constants can be found in Fig. S7 and Table S3). The colour scale (bottom) is scaled down to 5% of the original data above (notice changed scaling to $10^{-5}$). Details about the individual experiments (mass fraction, excitation energy) are given above the figures.

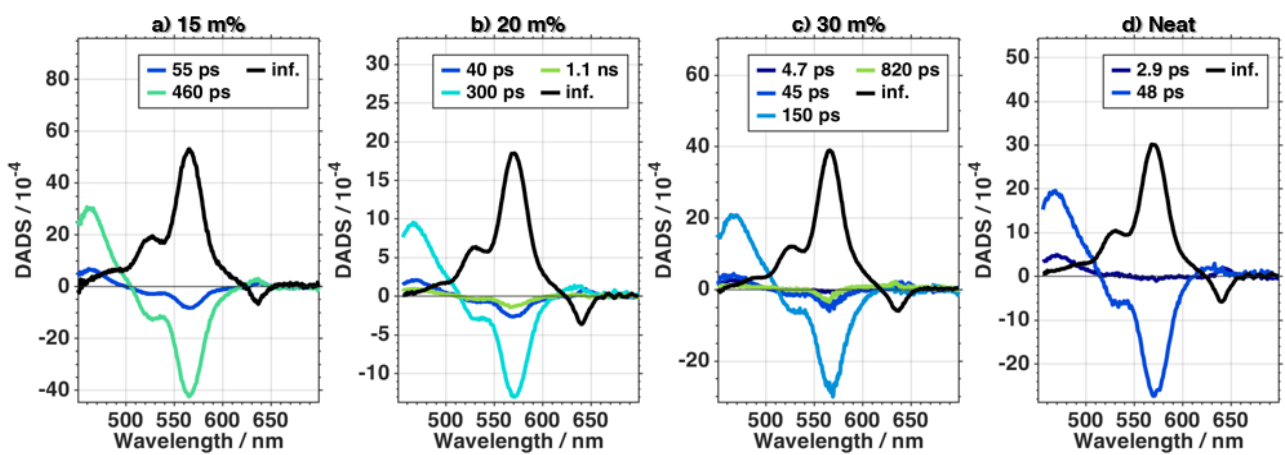

**Figure S10**. Decay-associated difference spectra (DADS) generated from the global analysis of results on **G-Pen** acquired with "short-time, Helios" TA setup. a)-d) DADS for experimental data shown in Figure S6.

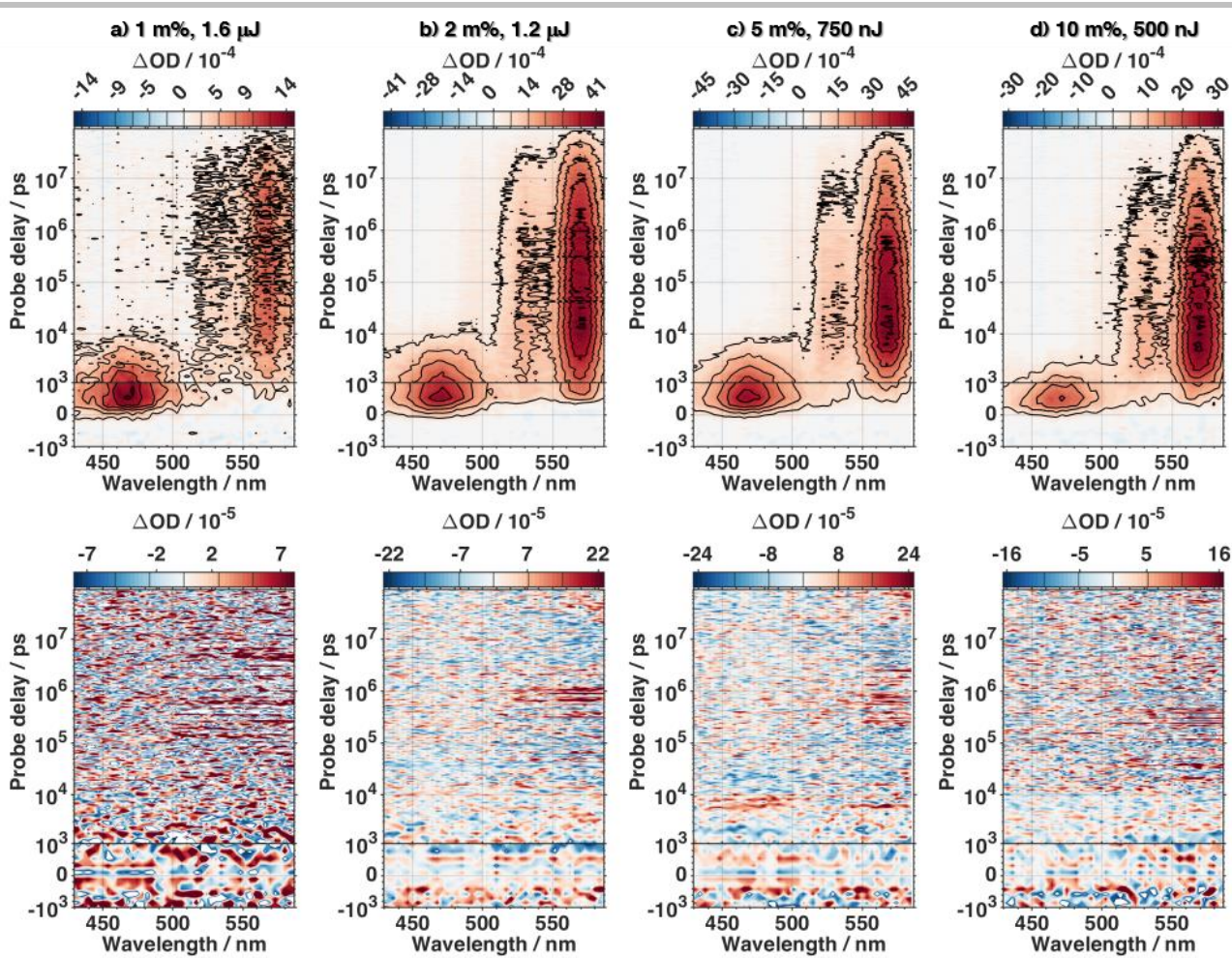

**Figure S11**. Evaluation of temporal evolution of transient absorption spectra of **G-Pen** in blends of PMMA (1-10 m%) measured with "long-time, EOS" TA setup. Experimental results for different mass fractions are shown in the top panels of a)-d). Black contours (top) are drawn at 15% intervals. The bottom panels show the residuals after subtraction of the fit (global analysis, time constants can be found in Fig. S9 and Table S3). The colour scale (bottom) is scaled down to 5% of the original data above (notice changed scaling to $10^{-5}$). Details about the individual experiments (mass fraction, excitation energy) are given above the figures.

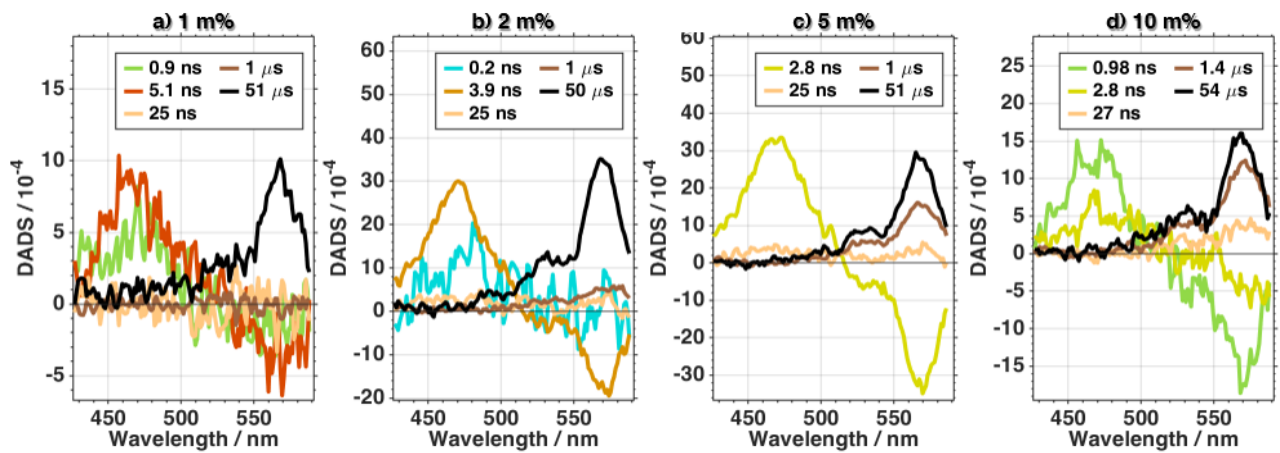

**Figure S12**. Decay-associated difference spectra (DADS) generated from the global analysis of results on **G-Pen** acquired with "long-time, EOS" TA setup. a)-d) DADS for experimental data shown in Figure S8.

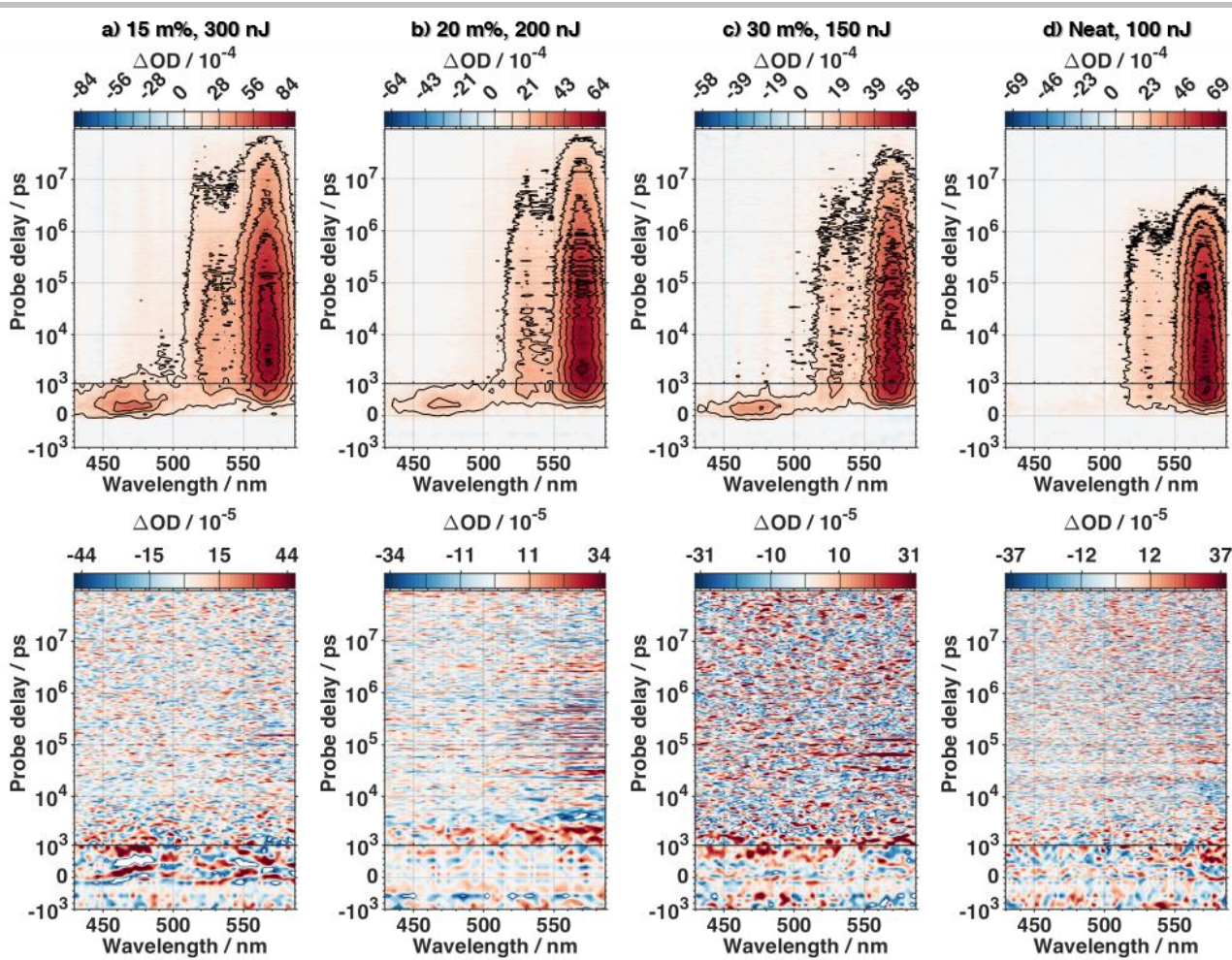

**Figure S13**. Evaluation of temporal evolution of transient absorption spectra of **G-Pen** in blends of PMMA (15-30 m%) and neat film measured with "long-time, EOS" TA setup. Experimental results for different mass fractions and the neat film are shown in the top panels of a)-d). Black contours (top) are drawn at 15% intervals. The bottom panels show the residuals after subtraction of the fit (global analysis, time constants can be found in Fig. S11 and Table S3). The colour scale (bottom) is scaled down to 5% of the original data above (notice changed scaling to $10^{-5}$). Details about the individual experiments (mass fraction, excitation energy) are given above the figures.

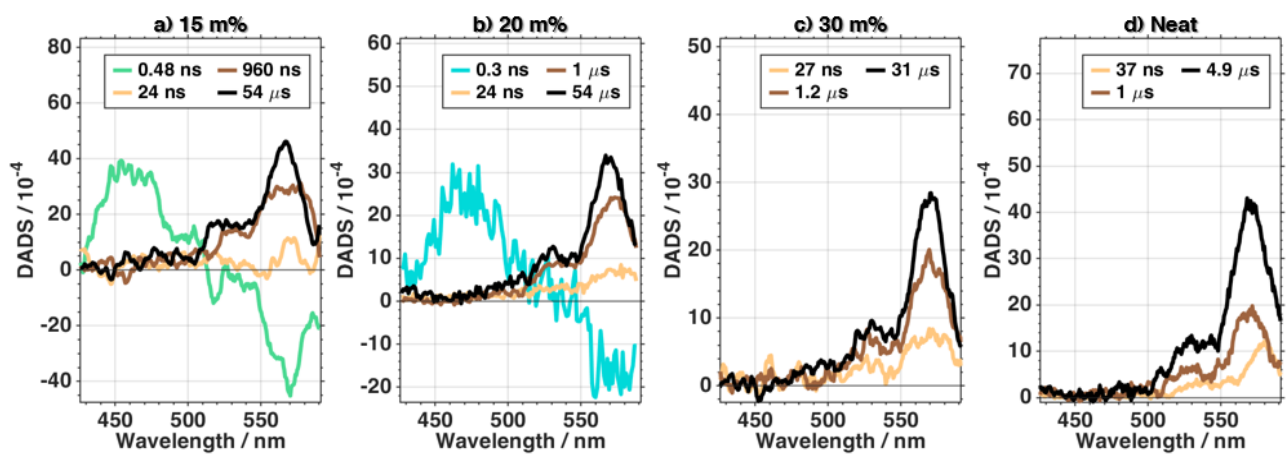

**Figure S14**. Decay-associated difference spectra (DADS) generated from the global analysis of results on **G-Pen** acquired with "long-time, EOS" TA setup. a)-d) DADS for experimental data shown in Figure S10.

## G-Pen – Results Summarized

**Table S4**: Summary of relevant time constants related to singlet-, and triplet decay pathways in **G-Pen**. $\tau_{S_1\rightarrow"T"}$ refers to the transition from the initial singlet ($S_1$) to triplet-related species: $T_1$ via ISC and $^1TT$ via SF. $\tau_{DC}$, $\tau_{TTA}$, and $\tau_{T_1}$ relate to the commonly observed bi-exponential decay pathways of the correlated triplet pair, which evolves over the separated $T_1T_1$ (decorrelation, DC) to the individual $T_1$'s (via triplet-triplet annihilation, TTA), before recovering to the ground-state via intersystem crossing. Calculation of the concentration and average molecular separation from the mass fraction (m%) of **G-Pen** in the PMMA blends is summarized in Table S2.

|  | $\tau_{S_1\rightarrow"T"}$ / ps | $\tau_{DC}$ / ns | $\tau_{TTA}$ / ns | $\tau_{T_1}$ / μs |
|---|---|---|---|---|
| neat thin film | 48 ± 1 | 37 ± 3 | 1.02 ± 0.09 | 4.9 ± 0.15 |
| G-Pen blend (PMMA) m% / % | | | | |
| 30 | 152 ± 7 | 27 ± 4 | 1.16 ± 0.08 | 31 ± 1 |
| 20 | 304 ± 15 | 24 ± 2 | 1.04 ± 0.07 | 54 ± 2 |
| 15 | 457 ± 42 | 24 ± 1 | 0.96 ± 0.1 | 54 ± 2 |
| 10 | 880 ± 76 | 27 ± 5 | 1.36 ± 0.12 | 54 ± 3 |
| 5 | 2440 ± 120 | 25 ± 8 | 1.0 ± 0.3 | 51 ± 1 |
| 2 | 4210 ± 255 | 25[a] | 1.0[a] | 50 ± 7 |
| 1 | 4930 ± 420 | 25[a] | 1.0[a] | 51 ± 10 |

[a] based on the results of other samples these time constants were kept as fixed components. As result, there is no error associated with these time constants.

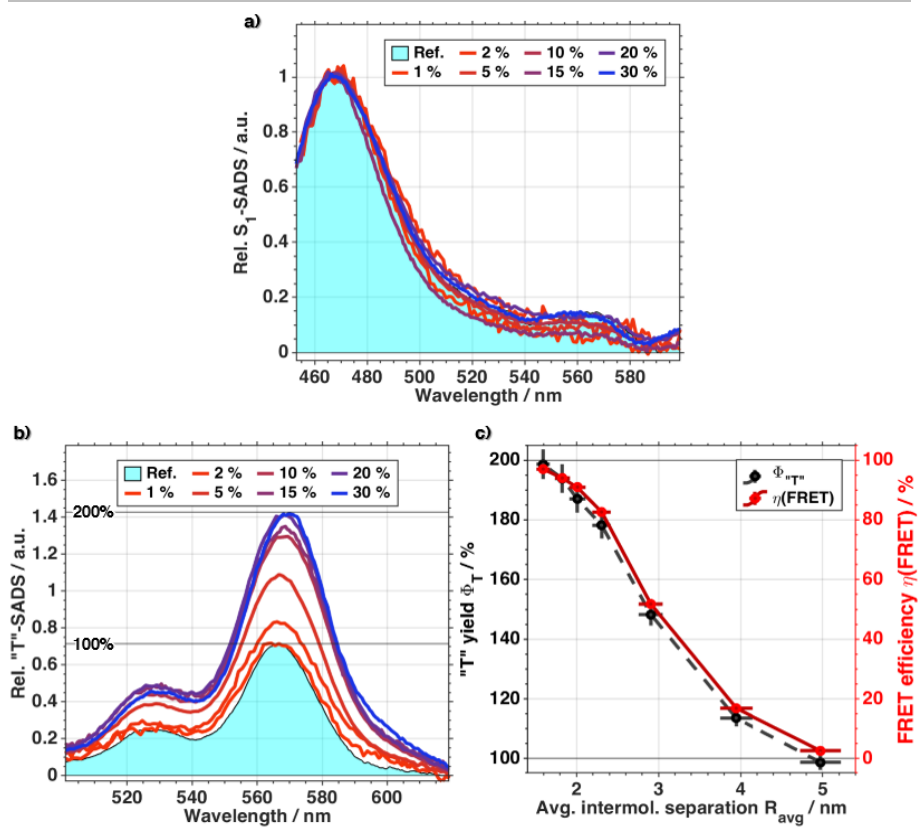

**Figure S15**. Summarized results for time-resolved experiments on **G-Pen**. a) and b) show the SADS of S1 and triplet-related ("T") spectra, normalized to the maximum of the S1-SADS. The SADS are generated after applying a sequential kinetic model on "short-time, Helios" experimental data with time constants shown in Figs. S5 and S7. The cyan area depicts the SADS obtained for THF measurements and is given as reference. In b) the T1-SADS of the reference was multiplied by a factor of (1/0.92, see Table 1) to account for the losses produced by the emission of **G-Pen**. The reference spectrum in b) constructs the reference (relative) extinction coefficient for "T"-spectra (can be $T_1$, $^{1,3,5}TT$ or $T_1+T_1$) from which the "T"-yield (black) shown in c) is calculated from. The FRET efficiency (red) in c) is calculated according to eq. 6 in the main text. Such correlation between "T"-yields and FRET efficiency is to be expected for the FRET-rate limited SF mechanism proposed in the main text. The larger deviation for large separations ($R_{avg}$) is to be expected, since the baseline QY for $T_1$ through ISC is 92% instead of 100%, reflecting the marginal yield of TT (<10%) at m% = 1% ($R_{avg}$ = 4.97 nm).

## TIPS-Pen – Experimental Data and Fits

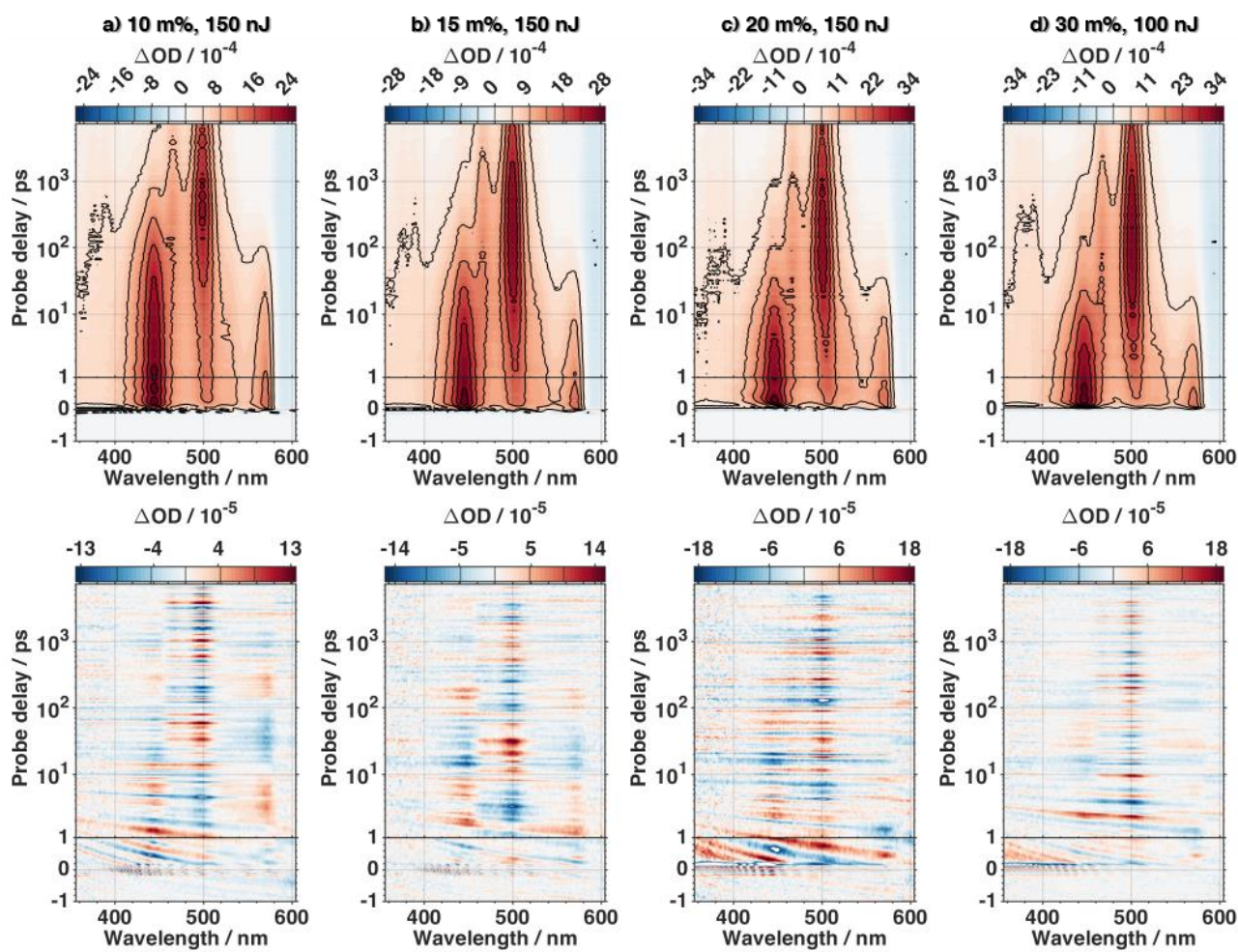

**Figure S16**. Evaluation of temporal evolution of transient absorption spectra of **TIPS-Pen** in blends of PMMA (10-30 m%) measured with "short-time, Helios" TA setup. Experimental results for different mass fractions are shown in the top panels of a)-d). Black contours (top) are drawn at 15% intervals. The bottom panels show the residuals after subtraction of the fit (global analysis, time constants can be found in Fig. S14 and Table S4). The colour scale (bottom) is scaled down to 5% of the original data above (notice changed scaling to $10^{-5}$). Details about the individual experiments (mass fraction, excitation energy) are given above the figures.

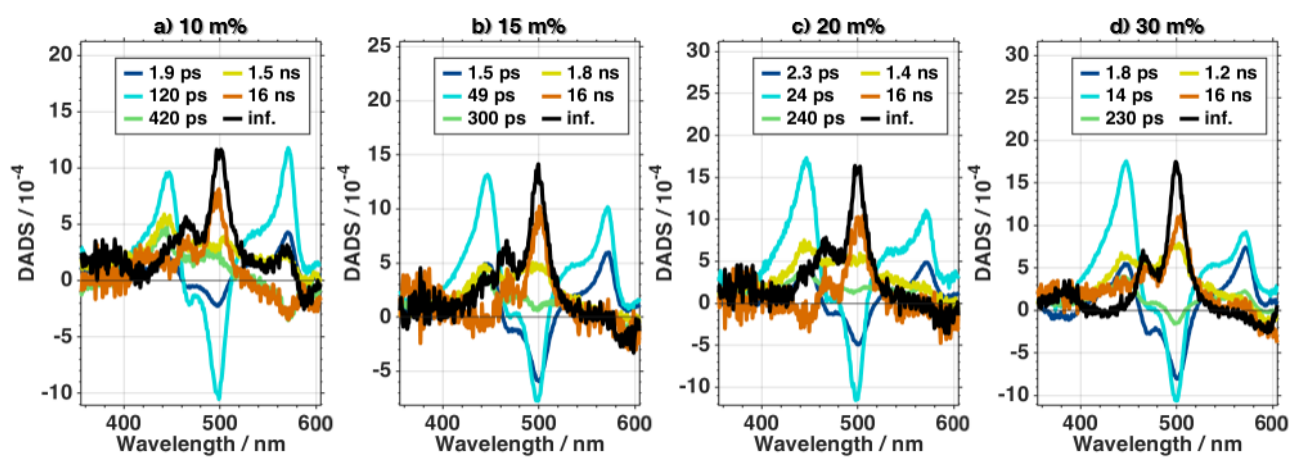

**Figure S17**. Decay-associated difference spectra (DADS) generated from the global analysis of results on **TIPS-Pen** acquired with "short-time, Helios" TA setup. a)-d) DADS for experimental data shown in Figure S13.

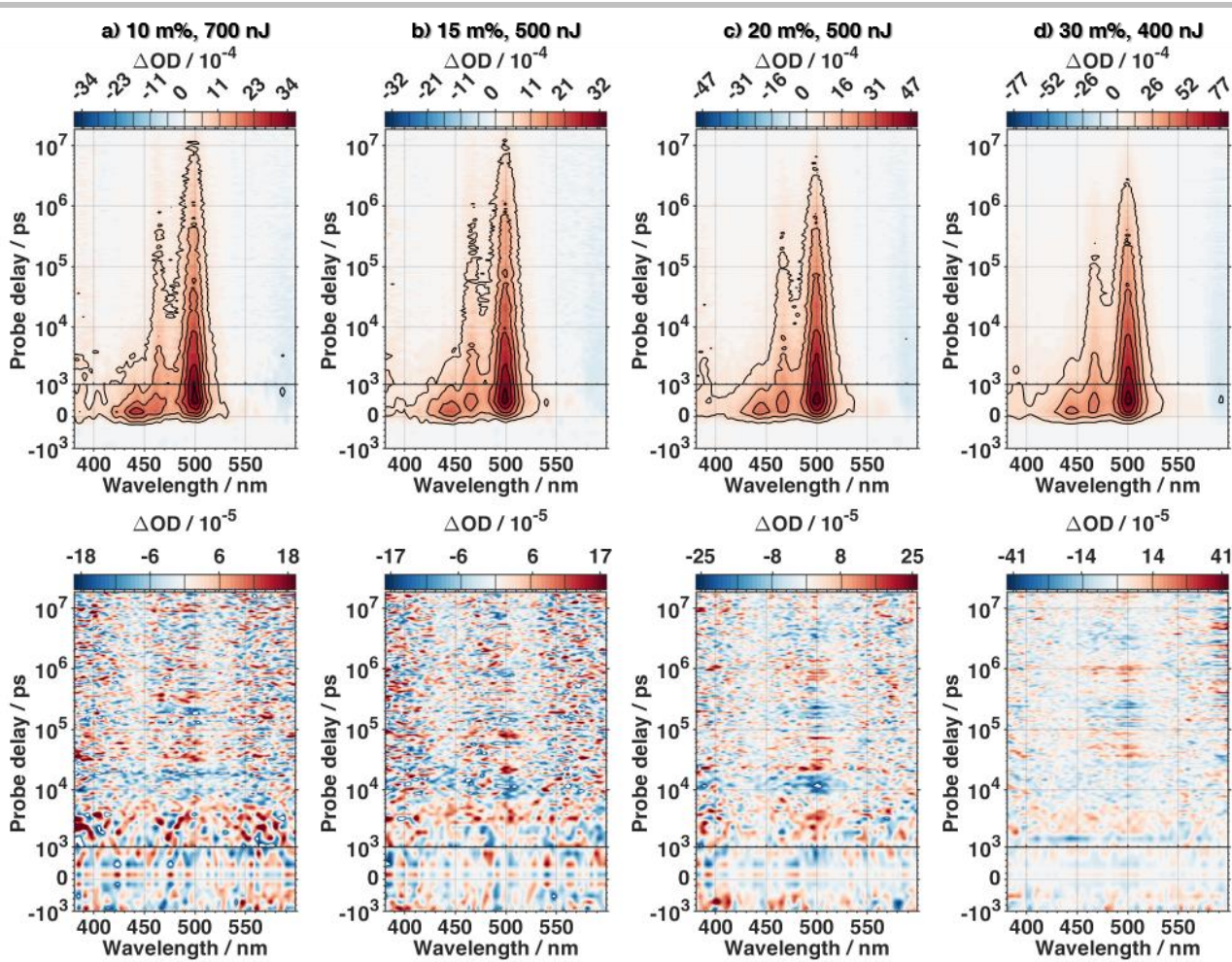

**Figure S18**. Evaluation of temporal evolution of transient absorption spectra of **TIPS-Pen** in blends of PMMA (10-30 m%) with "long-time, EOS" TA setup. Experimental results for different mass fractions and the neat film are shown in the top panels of a)-d). Black contours (top) are drawn at 15% intervals. The bottom panels show the residuals after subtraction of the fit (global analysis, time constants can be found in Fig. S16 and Table S4). The colour scale (bottom) is scaled down to 5% of the original data above (notice changed scaling to $10^{-5}$). Details about the individual experiments (mass fraction, excitation energy) are given above the figures.

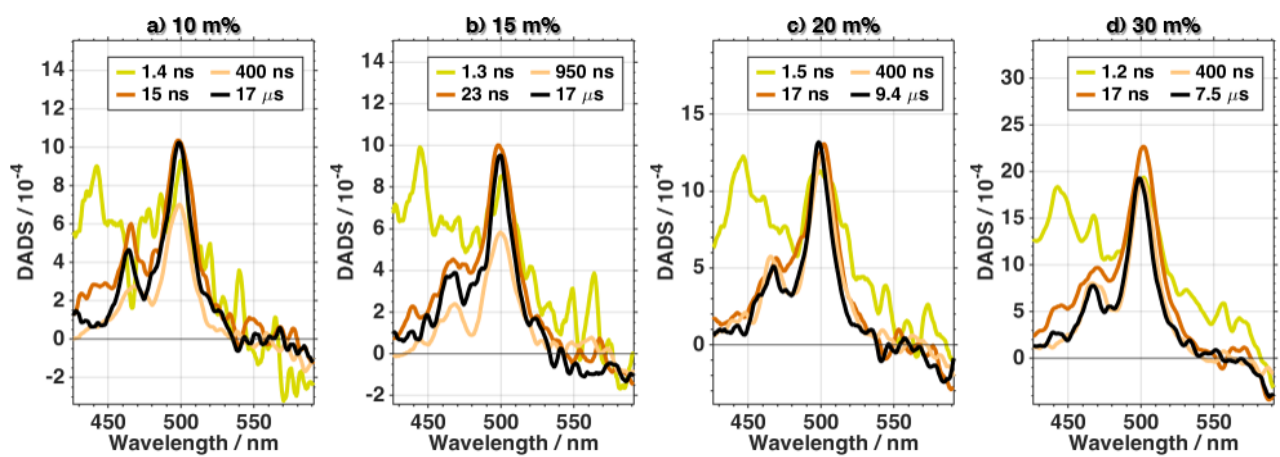

**Figure S19**. Decay-associated difference spectra (DADS) generated from the global analysis of results on **TIPS-Pen** acquired with "long-time, EOS" TA setup. a)-d) DADS for experimental data shown in Figure S15.

## TIPS-Pen – Results Summarized

**Table S5**: Summary of relevant time constants related to singlet-, and triplet decay pathways in **TIPS-Pen**. $\tau_{S_1 \to "T"}$ refers to the transition from the initial singlet ($S_1$) to triplet-related species: $T_1$ via ISC and CT-states + $^1TT$ via SF. $\tau_{DC}$, $\tau_{TTA}$, and $\tau_{T_1}$ relate to the commonly observed bi-exponential decay pathways of the correlated triplet pair, which evolves over the separated $T_1T_1$ (decorrelation, DC) to the individual $T_1$'s (via triplet-triplet annihilation, TTA), before recovering to the ground-state via intersystem crossing. Calculation of the concentration and average molecular separation from the mass fraction (m%) of **TIPS-Pen** in the PMMA blends is summarized in Table S2.

| TIPS-Pen blend (PMMA) m% / % | $\tau_{S_1 \to "T"}$ / ps | $\tau_{DC}$ / ns | $\tau_{TTA}$ / ns | $\tau_{T_1}$ / µs |
|---|---|---|---|---|
| 30 | 14.3 ± 1.8 | 17.1 ± 2.1 | 395 ± 45 | 7.5 ± 0.6 |
| 20 | 24.0 ± 2.4 | 16.5 ± 2.4 | 404 ± 60 | 9.4 ± 1.2 |
| 15 | 49 ± 8 | 23.3 ± 4.5 | 950 ± 200 | 16.7 ± 3.3 |
| 10 | 115 ± 16 | 15.3 ± 1.8 | 400 ± 70 | 16.6 ± 3.6 |

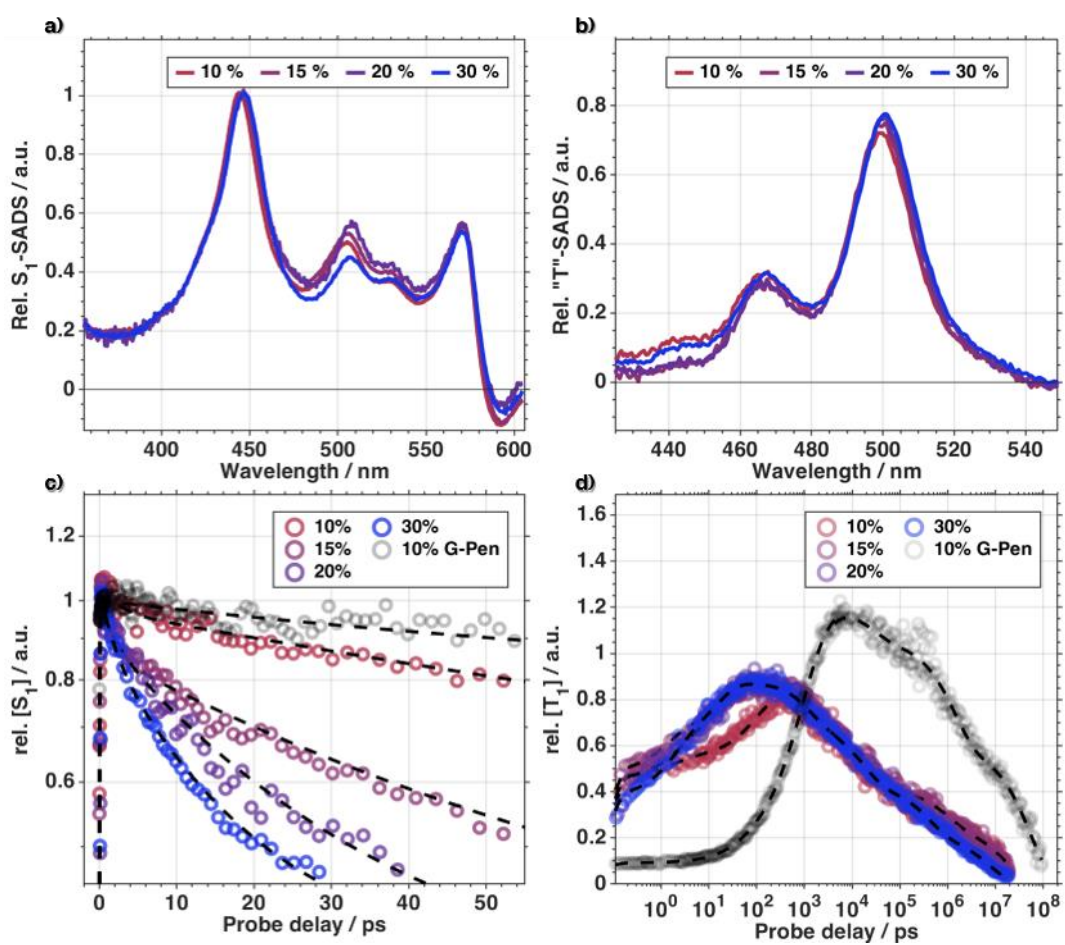

**Figure S20**. Summarized results for time-resolved experiments on **TIPS-Pen**. a) and b) show the SADS of S1 and triplet-related ("T") spectra, normalized to the maximum of the S1-SADS. The SADS are generated after applying a sequential kinetic model on "short-time, Helios" experimental data with time constants shown in Fig. S14. c) Normalized kinetic traces corresponding to the maximum of the $S_1$-ESA of **TIPS-Pen** (445 ± 10 nm) for different mass fractions in PMMA. For comparison, also a normalized trace of **G-Pen** (m% = 10%, 460 ± 10 nm) is plotted. d) Normalized (to maximum of $S_1$-SADS in a) kinetic traces corresponding to the maximum of the $^1TT$-ESA of **TIPS-Pen** (500 ± 5 nm) for different mass fractions in PMMA. For comparison, also a normalized trace of **G-Pen** (m% = 10%, 560 ± 10 nm) is plotted. The traces in d) consist of "short-time, Helios" (probe delays < 4*10³ ps) and "long-time, EOS" (probe delays > 4*10³ ps) to give a full picture of the triplet dynamics. The slower formation- and decay time of triplets can be discerned clearly by comparison of traces corresponding to **TIPS-Pen** and **G-Pen** in d). This is in agreement to what is expected from comparison of the time constants summarized in Tables S3 and S4 for **G-Pen** and **TIPS-Pen**, respectively.

# 6. Fitting Energy Transfer Models to Experimental Rates

The main finding of our study revolves around experimentally determined "singlet fission" ("SF")-rates in the diluted PMMA blends that follow energy transfer scaling. Especially a Förster-resonance energy transfer (FRET, eq. 3 in the main text) describes our results best. We want to scrutinize these findings by also considering a Dexter energy transfer (DET, eq. 2 in the main text) to fit our data. For this reason, we performed fits of experimental rate constants with the equations of the two models shown below:

$$k_{fit}^{FRET}(p) = \frac{1}{\tau_{S_1}} \cdot \left[1 + \left(\frac{\beta_1}{R(p)}\right)^6\right] \quad (S3)$$

$$k_{fit}^{DET}(p) = \frac{1}{\tau_{S_1}} + \beta_2 \cdot e^{\frac{-2 \cdot R(p)}{\beta_1}} \quad (S4)$$

For more clarity on which parameter varies during the fit, we replaced the physical parameters of $R_0$ (see eq. 3 in main text) in eq. S3 and L (see eq. 2 in main text) in eq. S4 with the fit-parameter $\beta_1$. The scaling factor $\beta_2$ in eq. S4 accounts for the pre-exponential term found in descriptions of the DET model.[13] An offset term ($\frac{1}{\tau_{S_1}} = k_{S_1}$) is added in both cases to account for the radiative- and non-radiative decay processes occurring at infinite (R = inf) separations. The offset in this case is not a fit parameter but is taken from the averaged results presented in section 4 of the SI (for **G-Pen**) and from results presented in literature (for **TIPS-Pen**)[34, 49], which is also summarized in Table 1 of the main text.

The goodness-of-fit ($\sigma_{fit}$) is determined by calculation of the R-squared value ($R^2$), which relates the variance of the sum-square residuals (experimental rates minus fitted rates ($k_{fit}$)) with the variance of the experimental rates, which is calculated from the difference of individual rates with the averaged experimental rates $\overline{k}_{exp}$. $R^2$ is calculated according to

$$R^2 = 1 - \frac{\sum_p \left(k_{exp}(p) - k_{fit}(p)\right)^2}{\sum_p \left(k_{exp}(p) - \overline{k}_{exp}\right)^2}, \quad (S5)$$

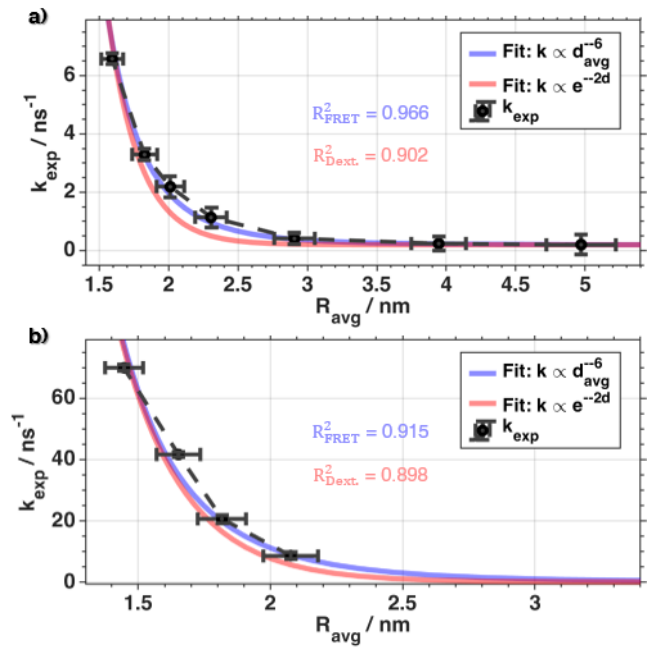

**Figure S21**. Fit of experimental "SF"-rates with energy transfer models. a) and b) show the fit of rates obtained for **G-Pen** and **TIPS-Pen**, respectively, with FRET (blue, eq. S3) and DET (red, eq. S4) models. The $R^2$ – values (eq. S5) are given in the plots.

resulting in optimum values of $R^2$ of 1 for a perfect fit of experiment observables. Though, the optimization algorithm tries to minimize the goodness-of-fit and it is calculated as $\sigma_{fit} = 1 - R^2$. Fit results using FRET (eq. S3) and DET (eq. S4) models on experimental rates obtained for **G-Pen** (p = 1-7, Table S3) and **TIPS-Pen** (p = 1-4, Table S4) are shown in Fig. S16a and S16b, respectively. Overall better fit quality is obtained with FRET in both cases. Especially in the case of **G-Pen** (Fig. S16a), a larger discrepancy between the models can be found. This is to be expected, as the short-range coupling parameter ($V_{short}$, eq. 1) which gives rise to the exponential scaling in DET (eq. S4) is more sensitive to intermolecular distances, leading to larger discrepancies at intermediate blend concentrations. This effect is not as pronounced in **TIPS-Pen** because of the accelerated kinetics.

To provide further arguments for the robustness of the FRET model in explaining experimental observations, we selected a subset of experimental rates to be fitted with the corresponding models. From eqs. S3-4, we can see that one and two fit parameters are required in the FRET and DET models, respectively. Conversely, a subset of two rate constants is sufficiently small to test our hypothesis. To be non-biased, we search all possible permutations of two rate constants ($k_{exp}(p)$, with p = {$p_1$, $p_2$}) to be representatives of the complete set, which is presented in Figs. S19 and S20 for **G-Pen** and **TIPS-Pen**, respectively. Points, where $p_1 = p_2$, fit one rate constant, instead of two (insufficient number of points for the DET model). Nevertheless, as can be seen in the comparison of $R^2$-valued for FRET (a) and DET (b) models, a generally better fit can be found for the former. The DET model only reaches appropriate resemblance (similar to fitting all points), when the largest rate constant ($p_{1,2}$ = {7, 1-7} in Fig. S19b or $p_{1,2}$ = {4, 1-4} in Fig. S20b) is considered in the fit. For the FRET model, the complete set of rate constants is sufficiently well fit by a larger number of subsets, even for those that consider just a single rate constant.

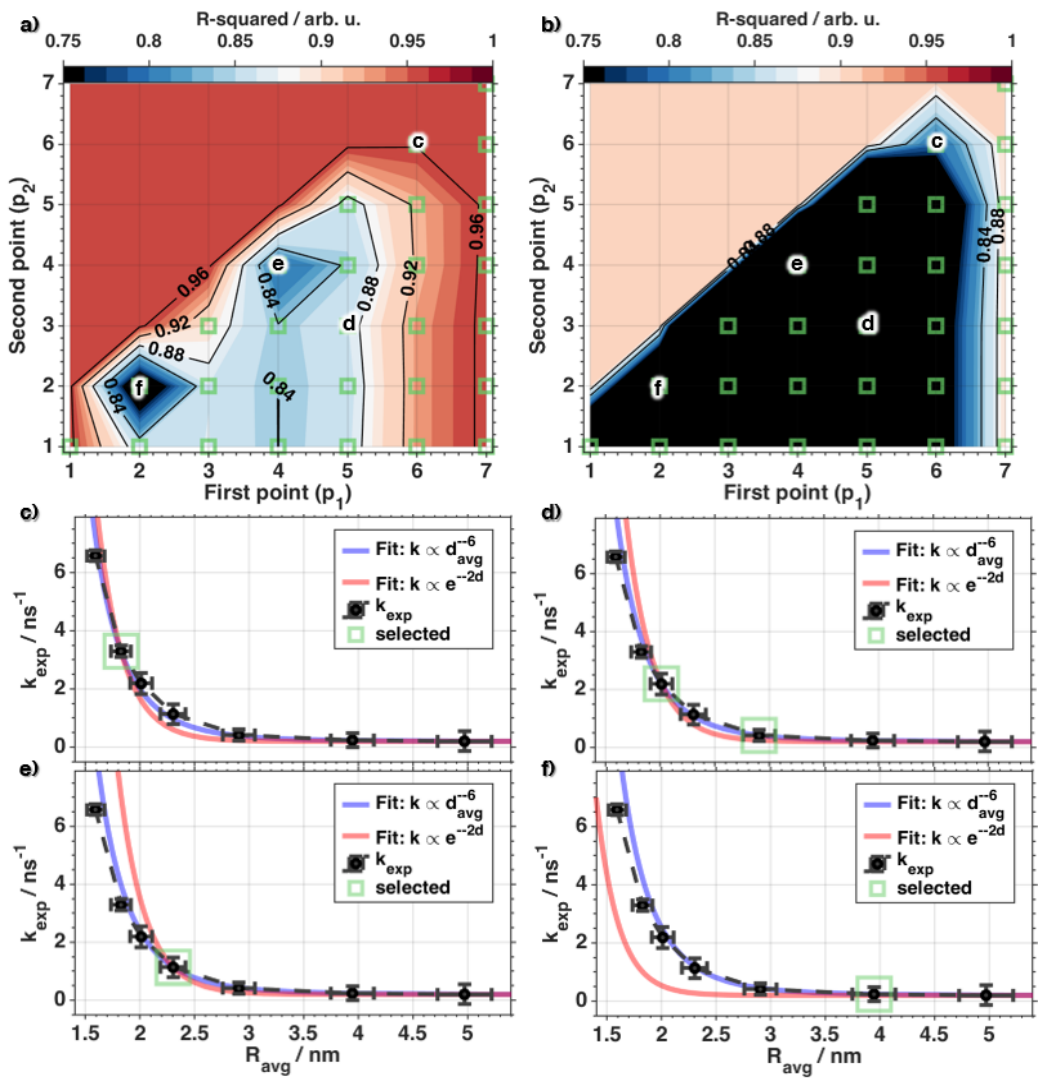

**Figure S22**. Exploring different rate constant subsets of **G-Pen** fit by FRET (a, eq. S3) and DET (b, eq. S4) models. Black contour levels are drawn in 0.04 intervals and coloured contours show 0.01 intervals. Green squares represent the different subsets of rates fitted {$p_1$,$p_2$}. Due to {$p_1$, $p_2$} and {$p_2$, $p_1$} being the same subset, only the lower diagonal is considered. The R-squared ($R^2$, eq. S5) value of the fit that considers all data points is plotted in the upper diagonal as reference value. c)-f) show the fit of the rate constants for the points indicated in a)-b).

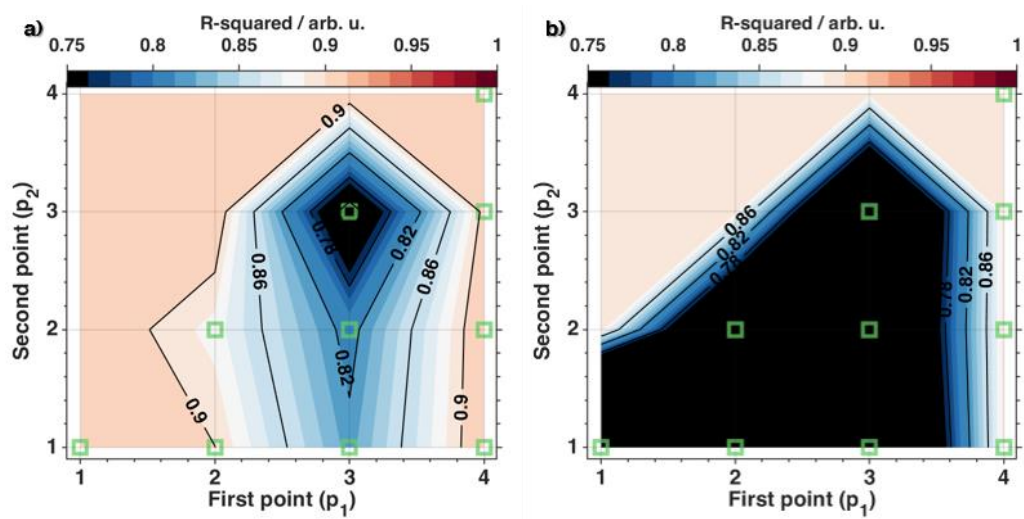

**Figure S23**. Exploring different rate constant subsets of **TIPS-Pen** fit by FRET (a, eq. S3) and DET (b, eq. S4) models. Black contour levels are drawn in 0.04 intervals and coloured contours show 0.01 intervals. Green squares represent the different subsets of rates fitted {$p_1$,$p_2$}. Due to {$p_1$, $p_2$} and {$p_2$, $p_1$} being the same subset, only the lower diagonal is considered. The R-squared ($R^2$, eq. S5) value of the fit that considers all data points is plotted in the upper diagonal as reference value.

# 7. Details on Calculation of FRET Rates

Details on the calculation of FRET rates ($k_{calc}$), based on eqs. 3-4 of the main text, are provided in the following. A thorough description of the procedure can be found in ref.[15]. The spectral overlap J (see Table 1 main text) is defined as

$$J = \int_0^\infty \varepsilon(\lambda) \cdot F_{em}(\lambda) \cdot \lambda^4 \, d\lambda, \tag{S6}$$

and is calculated from the overlap of absorption ($\varepsilon(\lambda)$) and area-normalized emission spectra ($F_{em}(\lambda)$) of the compounds, shown in Figure S21 for **G-Pen** and **TIPS-Pen**, respectively. The refractive indices ($n_x(m\%)$) of the blends in PMMA are estimated by mass-fraction weighed refractive indices of the compounds ($n_x$) and PMMA ($n_{PMMA}$)

$$n_x(m\%) = m\% \cdot n_x + (1 - m\%) \cdot n_{PMMA}. \tag{S7}$$

The refractive index of PMMA is taken from ref.[48] with a dispersion profile of

$$n_{PMMA}^2 = 2.1778 + (6.1209\lambda^2 - 1.5004\lambda^4 + 23.678\lambda^{-2} - 4.2137\lambda^{-4} + 7.3417 \cdot 10^{-1}\lambda^{-6} - 4.5042 \cdot 10^{-2}\lambda^{-8}) * 10^{-3} \tag{S8}$$

The refractive index of **G-Pen** and **TIPS-Pen** is calculated using Kramer-König relations,[53-54] that relates extinction coefficients (taken from ref. [1]) and refractive indices. An offset of $n_{TIPS-Pen}(\lambda_\infty) = 1.65$, retrieved from ref.[55], was applied. Since $n_{G-Pen}(\lambda_\infty)$ is unknown, it is estimated to be the same (1.65), considering the similarity of absorption profiles (the effect of varying $n_{G-Pen}(\lambda_\infty)$ from 1.5 to 1.8 is presented in Table S5). The effective refractive indices of the individual blends is calculated from the weighted amplitudes of the overlap spectrum (Figure S21a-b) multiplied with $n_x(m\%)$, which is summarized in Table S5. Excited-state lifetimes and emission QY's are summarized in Table 1 in the main text. Average intermolecular separations for **G-Pen** and **TIPS-Pen** are given in Table S2. For the orientation parameter kappa-squared ($\kappa^2$), a value of $\kappa^2 = 0.476$ is used, which is usually applied to static, but randomly oriented dipole moments of chromophores.[15] Which exact kappa-squared value is applicable to different scenarios is an extensively debated topic, which is outside the scope of this work and the reader is referred to ref. [56] for more details on this topic.

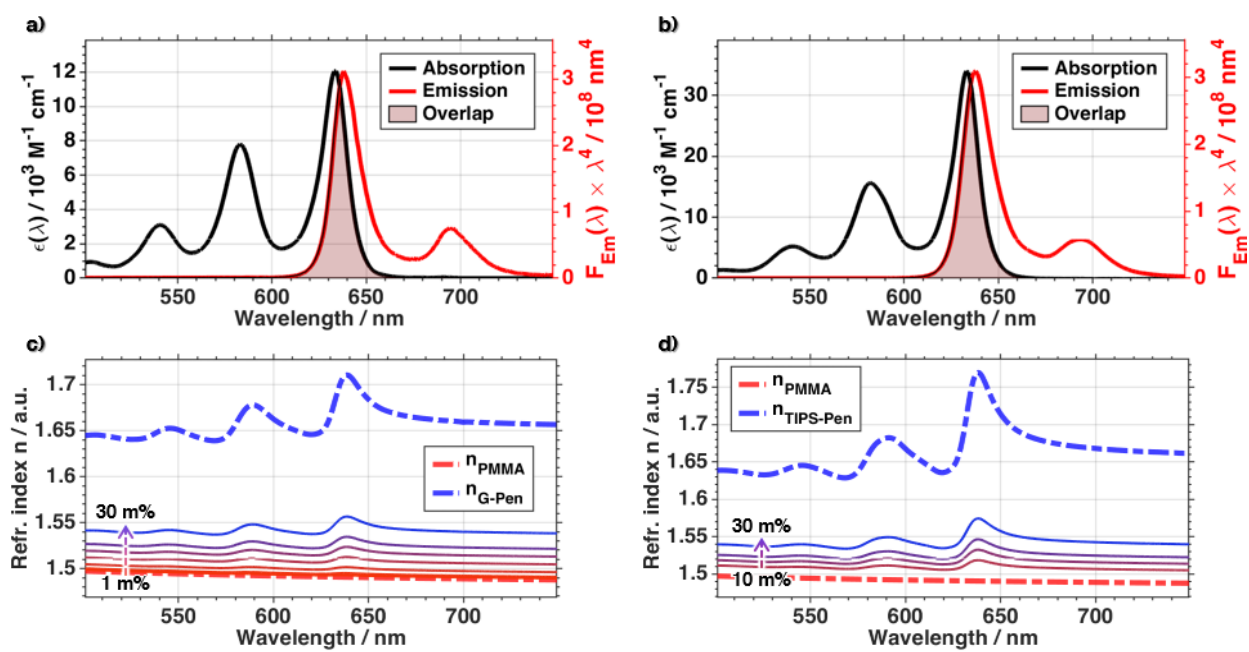

**Figure S24**. Relevant spectra for calculation of FRET rates. a) and b) Shows extinction and area-normalized emission spectra, multiplied by the fourth-power of the wavelength for **G-Pen** and **TIPS-Pen**, respectively. All spectra in a) and b) were taken from data provided in Ref.[1] (measured in hexane). The calculated overlap spectrum is also given as filled area. The integral of this spectrum represents J, which is given in Table 1 of the main text. c) and d) Presents refractive indices of **G-Pen** and **TIPS-Pen**, respectively. They were obtained by Kramers-König relation calculated from the extinction coefficients shown in a) and b). The refractive index of PMMA is calculated according to eq. S8. The continuous lines represent the weighted refractive indices of the blends (eq. S7) with various mass fractions (m%), ranging from 1-30 % in **G-Pen** and 10-30 % in **TIPS-Pen**.

**Table S6**. Summary of calculated refractive indices ($n_{eff}$, eqs. S7-S8) and FRET rate constants ($k_{calc}$, shown in Figure 3 of the main text) for **G-Pen** and **TIPS-Pen** according to equations 3-4 (and eq. S6). To the calculated rates, an offset value of $(\tau_{S1})^{-1}$ was added to account for radiative- and non-radiative decay processed at infinite separations, which is part of the experimental rates ($k_{exp}$). For **G-Pen**, a comparison of calculated FRET rates for different refractive indices is also provided, since exact value of $n_{G-Pen}(\lambda_\infty)$ is unknown. The refractive index has an effect at larger mass fraction (m%) and may lead to larger deviations from experimental rates for larger refractive indices, or in this case, can be understood as the standard deviation of the calculated rates.

| | G-Pen | | | | TIPS-Pen | | |
|---|---|---|---|---|---|---|---|
| **Mass fraction m% / %** | Effective refractive index $n_{eff}$ | Calculated rate $k_{calc}$ / (ps)$^{-1}$ | *Comparison of rates* $n_{G-Pen}(\lambda_\infty) = 1.50\ (1.80)$ / (ps)$^{-1}$ | Experimental rate $k_{exp}$ / (ps)$^{-1}$ | Effective refractive index $n_{eff}$ | Calculated rate $k_{calc}$ / (ps)$^{-1}$ | Experimental rate $k_{exp}$ / (ps)$^{-1}$ |
| **1** | 1.4927 | 0.205 | 0.205 (0.205) | 0.203 ± 0.017 | -- | -- | -- |
| **2** | 1.4948 | 0.226 | 0.226 (0.226) | 0.238 ± 0.014 | -- | -- | -- |
| **5** | 1.5009 | 0.371 | 0.375 (0.368) | 0.410 ± 0.020 | -- | -- | -- |
| **10** | 1.5110 | 0.877 | 0.905 (0.851) | 1.14 ± 0.098 | 1.5149 | 8.234 | 8.70 ± 2.42 |
| **15** | 1.5212 | 1.694 | 1.786 (1.608) | 2.19 ± 0.20 | 1.5270 | 17.728 | 20.6 ± 6.8 |
| **20** | 1.5314 | 2.800 | 3.014 (2.606) | 3.29 ± 0.16 | 1.5391 | 30.270 | 41.7 ± 8.3 |
| **30** | 1.5517 | 5.81 | 6.507 (5.200) | 6.58 ± 0.30 | 1.5634 | 63.035 | 69.9 ± 17.6 |